\documentclass[11pt]{article}
\usepackage{authblk}
\usepackage{amsmath}
\usepackage{url} 

\usepackage{geometry}
\usepackage{graphicx}
\usepackage{epstopdf}
\usepackage{
	amscd,amssymb,amsfonts,amsbsy,amsmath,amsthm,bm,booktabs,natbib,
	color,colortbl, caption,delarray, 
	enumerate,epsfig,epstopdf,extramarks,float,
	mathtools,multirow,natbib,xcolor
}
\usepackage{hyperref}
\hypersetup{
	colorlinks,
	linkcolor={red!50!black},
	citecolor={blue!50!black},
	urlcolor={blue!80!black}
}
\usepackage[shortlabels]{enumitem}

\definecolor{DarkBlue}{rgb}{0,.08,.45}
\definecolor{DarkRed}{rgb}{.7,0,.4}

\usepackage{titlesec}

\newcommand{\bpmt}{\begin{pmatrix}}
	\newcommand{\epmt}{\end{pmatrix}}
\newcommand{\ben}{\begin{enumerate}}
	\newcommand{\een}{\end{enumerate}}

\newcommand{\bea}{\begin{eqnarray*}}
	\newcommand{\eea}{\end{eqnarray*}}
\newcommand{\be}{\begin{eqnarray}}
\newcommand{\ee}{\end{eqnarray}}

\newcommand{\bsp}{\begin{split}}
	\newcommand{\esp}{\end{split}}
\newcommand{\ed}{\end{document}}

\newcommand{\btab}{\begin{tabular}}
\newcommand{\etab}{\end{tabular}}
\newcommand{\bc}{\begin{center}}
\newcommand{\ec}{\end{center}}

\newcommand{\bfi}{\begin{figure}}
\newcommand{\efi}{\end{figure}}

\newcommand{\bdes}{\begin{description}}
\newcommand{\edes}{\end{description}}
\newcommand{\bay}{\begin{array}}
\newcommand{\eay}{\end{array}}

\newcommand{\nn}{\nonumber}

\newtheorem{thm}{Theorem}
\newtheorem{lem}{Lemma}
\newtheorem{cor}{Corollary}

\newtheorem{prop}{Proposition}

\newcommand{\bass}{\begin{assumption}}
\newcommand{\eass}{\end{assumption}}
\newcommand{\bthm}{\begin{thm}}
\newcommand{\ethm}{\end{thm}}
\newcommand{\blem}{\begin{lem}}
\newcommand{\elem}{\end{lem}}
\newcommand{\bcor}{\begin{cor}}
\newcommand{\ecor}{\end{cor}}
\newcommand{\bpf}{\begin{proof}}
\newcommand{\epf}{\end{proof}}

\def\bco{\iffalse}
\def\cov{{\rm cov}}

\def\var{{\rm var}}

\def\diag{{\rm diag}}

\def\ci{\cite}
\def\cp{\citep}

\def\vvs{\vspace{0.1cm}}

\def\inv{^{-1}}

\def\tij{t_{ij}}
\def\Yij{Y_{ij}}

\def\i1{_{i1}}

\def\half{^{1/2}}

\DeclareMathOperator*{\argmin}{argmin}

\def\inv{^{-1}}

\def \mbb {\mathbb}
\def \mbf {\mathbf}
\def \mc {\mathcal}
\def \E {\mbb E}

\def\cT{\mathcal{T}}

\def\R{\mbb R}

\def\wh{\widehat}
\def\wt{\widetilde}
\def\lra{\longrightarrow}
\def\ra{\rightarrow}

\def\half{^{1/2}}

\def\s1n{\sum_{i=1}^n}
\def\p1n{\prod_{i=1}^n}
\def\i01{\int_0^1}

\def\references{\bibliography{merged}}

\def\tp{\top}
\def\E{E}
\def\R{\mbb{R}}
\def\cN{\mc{N}}
\def\cT{\mc{T}}
\def\diffop{\ \mathrm{d}}
\def\vvs{\vspace{.2cm}}
\def\cinD{\overset{\mc{D}}\lra}

\def\cv{{\rm CV}}
\def\fm {\frac 1 {m_i}}
\def\fn {\frac 1 n}
\def\half{\frac{1}{2}}

\def\mise {{\rm MISE}}
\def\oline{\overline}
\def\opt {{\rm opt}}

\def\pt{\partial}
\def\ptt{\frac{\pt}{\pt t}}
\def\pty{\frac{\pt}{\pt y}}
\def\p2t{\frac{\pt^2}{\pt t^2}}
\def\pysq{\frac{\pt^2}{\pt y^2}}
\def\si1n {\sum_{i=1}^n}
\def\sj1m {\sum_{j=1}^{m_i}}

\def\Yij{Y_{ij}}
\def\tij{t_{ij}}
\def\Ys{Y_i(s)}

\def\T{\mathcal{T}}

\title{Rank Dynamics for Functional Data}
\author[1]{Yaqing Chen\thanks{yaqchen@ucdavis.edu}}
\author[2]{Matthew Dawson}
\author[1]{Hans-Georg M{\"u}ller}
\affil[1]{Department of Statistics, University of California, Davis}
\affil[2]{Graduate Group in Biostatistics, University of California, Davis}
\date{May 7, 2020}

\begin{document}
\maketitle

\begin{abstract}
	The study of the dynamic behavior of cross-sectional ranks over time for functional data and the ranks of the observed curves at each time point and their temporal evolution can yield valuable insights into the time dynamics of functional data. This approach is of interest in various application areas. For the analysis of the dynamics of ranks, estimation of the cross-sectional ranks of functional data is a first step. Several statistics of interest for ranked functional data are proposed. To quantify the evolution of ranks over time, a model for rank derivatives is introduced, where rank dynamics are decomposed into two components. One component corresponds to population changes and the other to individual changes that both affect the rank trajectories of individuals. The joint asymptotic normality for suitable estimates of these two components is established. The proposed approaches are illustrated with simulations and three longitudinal data sets: Growth curves obtained from the Z\"urich Longitudinal Growth Study, monthly house price data in the US from 1996 to 2015 and Major League Baseball offensive data for the 2017 season.
\end{abstract}

\section{Introduction}\label{sec:intro}

In many statistical applications, practitioners are interested in relative, as opposed to absolute, behavior of random quantities. For example, 
in growth studies, one is often interested in growth faltering, stunting and more  generally determining whether children are tall, normal or small for their age. Such determinations  are based on an assessment of how individuals rank relative to others, where an individual's rank will change as the individual ages.  In sports, many interested parties aim to track the longitudinal changes in the relative rankings of the best players and teams. For example, the compensation a player receives is tied to relative performance. Related studies have been done on regression models for conditional distribution functions and quantiles \citep[][for example]{kim:07,kim:11,wu:13:4,tian:14,cho:16,kuru:18}, while our focus here is on modeling the temporal evolution of longitudinal ranks.

In the case of univariate measurements, ranking data is straightforward and well-studied. However,  one cannot rank multivariate data because  there is no total ordering in $\R^p$. For the same reason,  functional data that correspond to infinite-dimensional objects similarly cannot be ordered \ci[for overviews, see, e.g.,][]{rams:97,horv:12,wang:16}. In related work, the  analysis of sports data with functional data analysis techniques  has been recently considered by \citet{chen:18}, archetypoids of functional trajectories were applied to sports statistics by \citet{vinu:17}, and  \citet{mart:16:2} studied epigraph and hypograph indices which are the proportions of sample trajectories entirely lying above or below certain curves.

While functional data cannot be ordered, they are time-indexed and a total ordering exists cross-sectionally at each fixed time. This can be utilized to transform functional data into trajectories that consist of ranks, viewed as  functions of time. Of interest then is the modeling of the ranks of individuals and their patterns over time. 
In this paper, we discuss statistical tools to study such rank dynamics.  In particular, we introduce a novel decomposition for rank dynamics, where we show that  rank derivatives can be naturally decomposed into two components, corresponding to  a population and an individual contribution to the rank evolution, respectively. A simple example for the effect of the population on individual ranks occurs when the scores of the population improve overall, but a particular individual stays the same, say a runner maintains a certain level of speed but the population of runners at large is getting faster --- then the individual runner's rank will drop within the population, even though the individual's performance is not worse than before.

As rank dynamics depend on the interplay between individual and population changes and make reference to the cross-sectional population at each time $t$ where functional values are obtained, rank dynamics is quite  different from  common dynamic models in functional data analysis, where only the time dynamics of  individuals viewed by themselves are the focus, with the associated notions of
derivatives of observed trajectories and empirical dynamics. These previous approaches could be characterized as dynamics learning from functional data, and include derivative principal components, identification of differential equations, and
dynamic regression modeling
\cp{rams:02:2,wu:05,rams:07,wang:08:2, wang:08:3, hegl:09,mull:10:2,mull:18:6}.

More specifically, to study rank dynamics one first transforms  the observed functional data  through a probability transform that is implemented at each time point.  We assume that the functional data are densely sampled with negligible noise and that there is a stochastic process $Y$ with square integrable trajectories which are in the Hilbert space $L^2$. The process $Y$ generates the  sample of trajectories, which are the observed functional data.  If the functional data are measured on a  time grid with additive noise, one can implement a pre-smoothing step \cp{mull:06:11,hall:07:3}.  

Our starting point is the cross-sectional distribution 
\be\label{eq:cscdf} P(Y(t) \leq y)= F_t(y), \ee
for each $t \in \T$, where the domain $\T$ is a compact interval. Without loss of generality, we consider $\T = [0,1]$. The process of local probability transforms $R(t)$ associated with $Y$ is then 
\be  R(t)=F_t(Y(t)), \,\,  t\in  \T.  \nn
\ee
Since the subject-specific random process $R(t)$ conveys the information which fraction of individuals has larger and which fraction has lower values  at time $t$ compared to a  selected individual,  we refer  to $R(t)$ as the {\it rank process} associated with the functional process $Y$.

We note that the range of the rank process is always the interval $[0,1]$ and multiplying it by the sample size $n$ gives the actual ranks. Indeed, the distribution of $R(t)$ is uniform on $[0, 1]$ for every $t \in \mathcal{T}$, as it corresponds to the local probability transform. In a finite sample situation there are various ways to carry out the probability transform from a sample of data $Y(t)$,  depending on how one estimates the cumulative distribution function $F$. If one uses the empirical distribution function one obtains the actual ranks, but one can also use smooth versions of empirical distribution functions,  which often are advantageous \cp{falk:84} and yield approximate ranks.

The paper is organized as follows. In Section \ref{sec:decom}, we introduce a time-dynamic model for ranked functional data to quantify the temporal evolution of rank processes, which is a key contribution of this paper. 
In Section \ref{sec:sumstat}, we discuss several measures for the central tendency and variation of the rank trajectories, and in Section \ref{sec:est} the estimation of these population quantities. 
Asymptotic distributions  and finite-sample performance of the proposed estimates are demonstrated  in Sections \ref{sec:theory} and \ref{sec:simu}, respectively. Data illustrations are provided in  Section \ref{sec:app}, where we demonstrate rank dynamics for three scenarios including Z\"urich growth curves, house price trajectories and Major League Baseball data.

\section{A Time-Dynamic Model for Ranked Functional Data}\label{sec:decom}

Increases or decreases  in an individual's  rank trajectory depend on both the subject's functional trajectory $Y(t)$ and the functional trajectories of all other individuals in the sample, as the subject's rank at time $t$ depends on these two inputs.  This decomposition is exemplified by the  \emph{keeping up with the Joneses} paradigm,  where subjects' happiness is assessed through an individual's relative standing and its changes,   compared to their peers, i.e., critically important are the subject's rank and especially the changes in rank  \citep[e.g.,][]{barn:10,nguy:16}.

To quantify relative changes in a sample of functional data, it is expedient to utilize derivatives $R'(t)$. Recalling that $F_t(y)$ is the cross-sectional distribution of $Y$ at time $t$ and $R(t) = F_t(Y(t))$ and taking the derivative of $R$ with respect to $t$ leads to
\be
\aligned
R'(t)
&= C_1(t) + C_2(t)\\
&\coloneqq D_1(Y(t),t) + D_2(Y(t),t) Y'(t),\label{eq:rankDeriv}
\endaligned
\ee
where 
\be\label{eq:ptDeriv} D_1(y,t) \coloneqq \frac{\pt F_t(y)}{\pt t}\quad\text{ and }\quad D_2(y,t)\coloneqq \frac{\pt F_t(y)}{\pt y} = f_t(y).\ee

The two terms in \eqref{eq:rankDeriv} provide the decomposition of the rank derivative into two components for each subject. The first component $C_1(t)$ reflects the  changes in the distribution of the original process $Y$ with respect to time. More specifically, $C_1(t)$ indicates how population changes  influence the rank of a given  subject, where positive (negative) values of $C_1(t)$ for a specific  subject mean that the underlying functional trajectories $Y(t)$ for the other subjects are generally decreasing (increasing) at time $t$, which leads to an increase (decrease) in rank for the selected subject that is entirely due to a change in the characteristics of the general population.  On the other hand, the second component $C_2(t)$ represents the subject's own contribution to the rank dynamics.  Since $D_2(y,t) = f_t(y)\geq 0$, positive (negative) values of $Y'(t)$ contribute to  an increase (decrease) in rank due to individual change. Note that even if a subject's underlying functional trajectory $Y(t)$ is increasing, the population change $C_1(t)$ could increase even faster and  potentially overpower a subject's own contribution, leading to a decrease in rank.

To gain a better understanding of  the nature of the model in (\ref{eq:rankDeriv}), it is helpful to consider the case where $Y(t)$ is a constant function. In this case, we have that $C_2(t) = 0$ for all $t \in \mathcal{T}$, and the change in rank is completely determined by the rest of the population, i.e., the rank only changes when the population changes. Similarly, for a subject that traverses on a constant rank trajectory, it holds that  $R'(t) = 0$ for all $t$, which means that  population and subject driven components match each other, $C_1(t) = - C_2(t)$ for all $t$.

To determine the contributions of population and individual effects, it is then of interest to quantify the overall contributions of $C_1$ and $C_2$ to the rank derivative. For this, we define the rank component contributions 
\begin{equation}
\Lambda_1 \coloneqq \frac{\int_\cT \E(|C_1(t)|)\diffop t}{\int_\cT \E(|C_1(t)|)\diffop t + \int_\cT \E(|C_2(t)|)\diffop t}, \quad 
\Lambda_2 \coloneqq 1 - \Lambda_1.
\nn
\end{equation}
When $\Lambda_1$ is large, changes in rank are primarily dictated by changes in the population trajectories. In contrast, if $\Lambda_2$ dominates $\Lambda_1$, the changes in rank are due to changes in individual trajectories.


\section{Summary Measures for Rank Processes}\label{sec:sumstat}

Suppose we have a sample of trajectories $Y_i$ that are subject-specific independently and identically distributed realizations of a smooth underlying process $Y$, for $i=1,\dots,n$.  It is then of interest to have measures that quantify longitudinal central tendency and stability of both subject-specific and population ranks 
that are functionals of the corresponding rank processes $R_i(t)=F_t(Y_i(t))$ with $F_t$ as per \eqref{eq:cscdf} and $i= 1, \dots, n$.
A beneficial feature of the rank process approach is that like other rank-based methods, the analysis does not depend on the scale of the data and allows for direct comparisons of different data sources and measurement scales through comparing the corresponding rank processes.
\vspace{.2cm}

\paragraph{Subject-specific integrated rank} A natural way to summarize a subject's overall rank is to integrate the subject's rank trajectory over the time domain, i.e., to consider the subject-specific measure
\begin{equation}
\rho_i \coloneqq \int_\mathcal{T} R_i(t)\diffop t.
\label{eq:intRank}
\end{equation}

\paragraph{Subject-specific rank volatility} It is also of interest to quantify how variable a subject is in terms of rank,
which can be quantified by
\begin{equation}
\nu_i \coloneqq \int_\mathcal{T} [R_i(t) - \rho_i]^2\diffop t. \label{eq:intRankVar}
\end{equation}

\paragraph{Subject-specific rank dynamics} For smooth rank processes, one can define a rank derivative $R'(t)$, $t \in \mathcal{T}$. If  it is non-zero, then the subject's rank trajectory crosses the trajectories of other subjects, i.e., the rank of the subject will change over time. Pertinent  measures include
\begin{equation}
\zeta_i \coloneqq \int_\mathcal{T} R_i'(t)\diffop t = R_i(1) - R_i(0), \quad \text{and}\quad \eta_i \coloneqq \int_\mathcal{T} R_i'^2(t)\diffop t, \label{eq:mixing}
\end{equation}
quantifying  how variable the rank of a subject is over the time interval. \vspace{.2cm}

\paragraph{Population rank stability} 
Since  $\E[R(t)] = 1/2$ for all $t \in [0,1]$,  we have that $\E[R'(t)]=0$ under mild assumptions.  Although the mean functions are therefore not interesting, the variation of $R'$ on subdomains is of interest, as it  can  pinpoint temporal regions  where ranks tend to change and  the intensity of pairwise crossings of the rank trajectories is high. We define time-dependent rank stability as 
\begin{equation}
\gamma(t) \coloneqq \var[R'(t)] = \E[R'(t)^2].
\label{eq:timeMixing}
\end{equation}
Integrating this quantity leads to an overall population  rank stability coefficient, for which we choose
\begin{equation}
G \coloneqq  \exp{\left(-\int_\mathcal{T} \gamma(t)dt\right)}.
\label{eq:overallMixing}
\end{equation}
Note that if the underlying functional data never cross paths, then $\gamma(t) = 0$ for all $t$, and thus the overall rank stability is $G=1$, while the closer $G$  is to 0,  the lower is rank stability, i.e.,  the trajectories of the functional data exhibit more frequent crossings.

\section{Estimation}\label{sec:est}

The starting point is to estimate the rank trajectories $R_i(t)$.
Suppose for all subjects, processes $Y_i$ are observed on a regular dense grid $t_{i1}< \cdots< t_{im_i}$ on the time domain, i.e.,  there exists a design distribution function $\theta:\cT\ra[0,1]$ such that $t_{ij} = \theta\inv((j-1)/(m_i-1))$ for $j=1,\dots,m_i$ and $\Yij= Y_i(\tij)$. 
We assume that the underlying surface $F_t(y) = P(Y(t) \le y)$ is differentiable in both $y$ and $t$.
To obtain smooth estimates of the rank process, we utilize  a kernel function $K$, which is a pdf, and an integrated kernel $H$, which is a cdf.
Furthermore, we assume:

\ben[label=(A\arabic*)]
\item With probability 1, the process $Y$ has continuously differentiable sample paths and there exists a constant $M>0$ such that $\sup_{t\in\cT}|Y'(t)| \le M$.\label{ass:a0}
\item  \label{ass:a1} The kernel $K$ is a symmetric pdf on $\R$ such that,
\[\int x^l K(x)\diffop x <\infty,\text{ for } l=2,4.\]
The kernel $H$ is a cdf such that its derivative $H'(\cdot)$ exists almost everywhere, is bounded on $\R$ and is a symmetric pdf such that
\[\int x^l H'(x)\diffop x <\infty,\text{ for } l=2,4.\]
\item The kernel $K$ has a compact support, assumed to be $[-1,1]$. On $(-1,1)$, the first and second derivatives $K'$ and $K''$ exist and are bounded.\label{ass:a2}
\item The design distribution function $\theta$ is four times continuously differentiable on $[0,1]$. There exist $0<a_1<a_2$ such that $a_1\le \theta'(t)\le a_2$ for all $t\in[0,1]$.\label{ass:a3}
\een

We provide two strategies for the estimation of $R(t)$ based on the sample $\{t_{ij}, Y_{ij}\}$ as follows. 

\vvs

\paragraph{Cross-sectional empirical distributions} The most straightforward approach to obtain a ranked sample from a dense functional sample is to estimate the empirical distribution at each time point $t \in \mathcal{T}$. Obtaining cross-sectional empirical distributions in this manner is equivalent to taking cross-sectional ranks and scaling them, i.e., 
\begin{equation}
\wh R_i(t) = \frac{1}{n}\sum_{l \neq i} \mbf{1}_{\{Y_l(t) \leq Y_i(t)\}}.
\label{eq:empRank}
\end{equation}
The empirical ranking defined in \eqref{eq:empRank} has several benefits. It is very simple to implement, and its interpretation is very clear. However, since we aim to obtain differentiable rank functions that allow us to study the decomposition of rank dynamics into population and individual components, we need smooth estimates of the rank processes.\vvs 

\paragraph{Smooth rank functions} Smooth estimation of conditional/cross-sectional distribution functions has been well investigated \citep[e.g.,][]{hall:99:2,wu:13:4,vera:14,bela:17}. Define
\begin{align*}
\wt Q_{1i}(y,t) &=
\fm \sj1m h_T^{-1} H\left(\frac{y-\Yij}{h_Y}\right) K\left(\frac{t- \tij}{h_T}\right),\\
\wt Q_{2i}(y,t) &= \fm \sj1m h_T^{-1} K\left(\frac{t-\tij}{h_T}\right),
\end{align*}
and for $l=1,2,$
\[
\oline Q_l(y,t) = \fn \si1n \wt Q_{li}(y,t) ,
\]
where $h_Y,h_T>0$ are bandwidths.
Here, we utilize a kernel estimate of $F_t(y)$ given by well-established methods described in \citet{rous:69} and \citet{sama:89},
\be\label{eq:estF}
\wt F_t(y) = \frac{\oline Q_1(y,t)} {\oline Q_2(y,t)}.
\ee
Thus, a smooth estimator for the rank process $R_i(t)$ can be obtained by
\be
\wt R_i(t) = \wt F_t(Y_i(t)).
\label{eq:smoothRank}
\ee
We will discuss the selection of bandwidths $h_Y$ and $h_T$ in the Supplementary Material.

Using one of the two methods described above, we obtain the estimated rank for level $\Yij$ at time $\tij$, yielding the surface $\{\tij, \Yij, \wh{R}_i(\tij)\}$ or $\{\tij, \Yij, \wt{R}_i(\tij)\}$, and hence  estimate the measures $\rho_i$, $\nu_i$ and $\zeta_i$ given in \eqref{eq:intRank}--\eqref{eq:mixing}, respectively, by plugging in either of the two estimators of $R_i(t)$, applying numerical  integration. Estimation of the measures $\eta_i$, $\gamma(t)$ and $G$ \eqref{eq:mixing}--\eqref{eq:overallMixing} requires  the estimation of the rank derivatives $R'(t)$, while 
identifying the components of the time-dynamic model as per \eqref{eq:rankDeriv} requires estimation of  $D_1(y,t)$, $D_2(y,t)$, and $Y'(t)$. 

For estimating $Y'(t)$, one can make use of local polynomial smoothing, or a similar method.  
To estimate $D_1(y,t)$ and $D_2(y,t)$ defined in \eqref{eq:ptDeriv}, we take partial derivatives of \eqref{eq:estF}, yielding
\be\label{eq:estD}
\wt D_1(y,t)
=\frac{\oline Q_3(y,t)} {\oline Q_2(y,t)} - \frac{\oline Q_1(y,t)\oline Q_4(y,t)}{\oline Q_2(y,t)^2} \text{\quad and \quad}
\wt D_2(y,t)
= \frac{\oline Q_5(y,t)} {\oline Q_2(y,t)},
\ee
where
\begin{align*}
\wt Q_{3i}(y,t) &= \fm \sj1m h_T^{-2}H\left(\frac{y- \Yij}{h_Y}\right)K'\left(\frac{t-\tij}{h_T}\right), \\
\wt Q_{4i}(y,t) &=\fm \sj1m h_T^{-2} K'\left(\frac{t -\tij}{h_T}\right),\\
\wt Q_{5i}(y,t) &= \fm \sj1m h_Y^{-1}h_T^{-1} K\left(\frac{y- \Yij}{h_Y}\right) K\left(\frac{t-\tij}{h_T}\right), 
\end{align*}
and for $l=3,4,5$,
\[
\oline Q_l(y,t) = \fn \si1n \wt Q_{li}(y,t),
\]
where $h_Y,h_T>0$ are bandwidths as in $\wt Q_{1i}$ and $\wt Q_{2i}$.

For subject $i$, the estimated components are 
\be\label{eq:rankCompEst} \wt C_{1i}(t) = \wt D_1(Y_i(t),t)\quad\text{ and }\quad\wt C_{2i}(t) = \wt D_2(Y_i(t),t) \wt Y_i'(t), \ee
where $\wt Y_i'(t)$ is an estimate of the derivative for example by local polynomial smoothing. From these estimators we obtain the estimated decomposition $\wt{R}'_i(t) = \wt{C}_{1i}(t) + \wt{C}_{2i}(t).$ 
The component contributions $\Lambda_1$ and $\Lambda_2$ may be estimated by numerically integrating the estimated components $\wt{C}_{1i}(t)$ and $\wt{C}_{2i}(t)$, 
\begin{equation}
\wt{\Lambda}_1 = \frac{\int_\mathcal{T} n^{-1}\sum_{i=1}^n|\wt{C}_{1i}(t)|\diffop t}{\int_\mathcal{T} n^{-1}\sum_{i=1}^n|\wt{C}_{1i}(t)|\diffop t + \int_\mathcal{T} n^{-1}\sum_{i=1}^n|\wt{C}_{2i}(t)|\diffop t}, \quad \text{and} \quad
\wt{\Lambda}_2 = 1 - \wt{\Lambda}_1.
\label{eq:rankContriEst}
\end{equation}
The measures $\eta_i$ in \eqref{eq:mixing} can then be estimated by plugging in $\wt{R}'_i(t)$ based on trajectory $Y_i(t)$; estimators for $\gamma(t)$ and $G$ in \eqref{eq:timeMixing} and \eqref{eq:overallMixing} are  obtained using the sample mean of $\wt R'_i(t)^2$.

\section{Theoretical Justifications}\label{sec:theory}
We demonstrate the asymptotic normality of $\wt F_t(y)$, the joint asymptotic normality of $[\wt D_1(y(t),t),\newline\wt D_2(y(t),t) y'(t)]^\tp$, given a curve $y(t)$, and the asymptotic normality of $\wt R'(t) = \wt D_1(y(t),t) + \wt D_2(y(t),t) y'(t)$. We denote convergence in distribution by $\cinD$, and define 
$$\sigma^2(K) = \int x^2K(x)\diffop x, \quad\text{and}\quad \sigma^2(H') = \int x^2H'(x)\diffop x.$$
All proofs and auxiliary results are in the Supplementary Material. Throughout, we use the notations $F_{s,s'}(z,z')= P(Y(s)\le z, Y(s')\le z')$ and $f_{s,s'}(z,z')$ for the joint cdf and pdf of $Y(s)$ and $Y(s')$,
and also the notation $\sim$, where $h_n\sim n^\alpha$ indicates $\lim_{n\ra \infty} h_n n^{-\alpha} = 1$. We further need to assume:

\ben[label=(A\arabic*),resume]
\item The partial derivatives $\frac{\pt^{k+l}}{\pt t^k\pt y^l} F_t(y)$ are bounded over $t\in[0,1]$ and $y\in\mbb R$, for $(k,l)\in \{(3,0),(0,3),(2,1),(1,2)\}$.\label{ass:a4}
\item The partial derivatives $\frac{\pt^2 }{\pt s\pt s'}F_{s,s'}(z,z')$, $\frac{\pt^2 }{\pt z\pt z'}F_{s,s'}(z,z')$ and $\frac{\pt^2 }{\pt s\pt z'}F_{s,s'}(z,z')$ are bounded over $s,s'\in[0,1]$ and $z,z'\in\R$.\label{ass:a5}
\een

The following proposition is similar to some results in literature, for example \citet{rous:69}; we omit the proof. Theorem \ref{thm:ANofD} is our main result. The proof and auxiliary lemmas are in the Supplementary Material. 

\begin{prop}\label{thm:ANofF}
	Assume \ref{ass:a0}--\ref{ass:a4},
	optimal bandwidth sequences  $h_Y\sim n^{-1/4}$ and $h_T\sim n^{-1/4}$, as $n,m_i\ra\infty$ with $\lim_{n\ra \infty} \max_{1\le i\le n}m_i\inv n^{1/2} = 0$. Then the estimate for $F_t(y)$ as defined in \eqref{eq:estF} satisfies 
	\be\nn
	\sqrt n \left[\wt F_t(y)-F_t(y)\right] \cinD  \cN \left( \beta_{\wt F}, \sigma_{\wt F}^2 \right), 
	\ee
	where
	\begin{gather*}
	\beta_{\wt F} = \half\sigma^2(H')\pty f_t(y) + \half \sigma^2(K)\left[\p2t F_t(y) + 2\frac{\theta''(t)}{\theta'(t)} \ptt F_t(y)\right],\\
	\text{and}\quad \sigma_{\wt F}^2  = F_{t,t}(y,y) - F_t(y)^2.
	\end{gather*}
\end{prop}
\bthm \label{thm:ANofD}
Assume \ref{ass:a0}--\ref{ass:a5}. 
Given a curve $y(t)$, the estimates $\wt C_1(t) = \wt D_1(y(t),t)$, $\wt C_2(t) = \wt D_2(y(t),t) y'(t)$ with $\wt D_1$ and $\wt D_2$ defined in \eqref{eq:estD} for the two components $C_1(t) = D_1(y(t),t)$
and $C_2(t) = D_2(y(t),t)y'(t)$	with $D_1$ and $D_2$ as per \eqref{eq:ptDeriv} are jointly asympotically normal. 
With bandwidths $h_Y\sim n^{-1/4}$ and $h_T\sim n^{-1/4}$, as $n,m_i\ra\infty$ such that $\lim_{n\ra \infty} \max_{1\le i\le n}m_i\inv n^{3/4} = 0$,
\[\sqrt n \left[ \bpmt \wt C_1(t)\\ \wt C_2(t)\epmt - \bpmt C_1(t)\\ C_2(t)\epmt \right] \cinD \cN\left(\bm\beta_{\wt C}, \bm\Sigma_{\wt C}\right),\]
where
\[\bm\beta_{\wt C} =\bpmt \half\sigma^2(H') \frac{\pt^2}{\pt t\pt y}f_t(y(t)) + \half\sigma^2(K) \left[\frac{\pt^3}{\pt t^3}F_t(y(t))  + 2\ptt \frac{\theta''(t)\ptt F_t(y(t)) }{\theta'(t)} \right]\\
\half\sigma^2(K)y'(t) \left[\frac{\pt^2}{\pt y^2}f_t(y(t)) + \p2t f_t(y(t)) + 2\frac{ \theta''(t)\ptt f_t(y(t))}{\theta'(t)}\right]\epmt,\]
and
\[\bm\Sigma_{\wt C} = \bpmt \Sigma_{11} & \Sigma_{12}\\ \Sigma_{12} &  \Sigma_{22} \epmt,\]
with
\begin{align*}
\Sigma_{11} &= \frac{\pt^2}{\pt s\pt s'}F_{t,t}(y(t),y(t)) - \left[\ptt F_t(y(t))\right]^2,\\
\Sigma_{12} &= y'(t) \left[\frac{\pt^2}{\pt s\pt z'}F_{t,t}(y(t),y(t)) - f_t(y(t))\ptt F_t(y(t))\right],\\
\Sigma_{22} &= y'(t)^2 \left[f_{t,t}(y(t),y(t)) - f_t(y(t))^2\right].
\end{align*}
\ethm
By continuous mapping,  the asymptotic normality of $\wt R'(t) = \wt C_1(t) + \wt C_2(t)$ follows.
\begin{cor}
	Under the assumptions of Theorem \ref{thm:ANofD}, with $R'(t) = C_1(t) + C_2(t)$, 
	\[\sqrt n \left[ \wt R'(t) - R'(t)\right] \cinD \cN \left(\beta(t), \sigma^2(t)\right), \]
	where 
	\begin{align*}
	\beta(t) &= \half\sigma^2(H') \frac{\pt^2}{\pt t\pt y}f_t(y(t)) + \half\sigma^2(K) \left[\frac{\pt^3}{\pt t^3}F_t(y(t))  + 2\ptt \frac{ \theta''(t)\ptt F_t(y(t))}{\theta'(t)} \right]\\
	&\quad + \half\sigma^2(K)y'(t) \left[\frac{\pt^2}{\pt y^2}f_t(y(t)) + \p2t f_t(y(t)) + 2\frac{ \theta''(t) \ptt f_t(y(t))}{\theta'(t)}\right],
	\end{align*}
	and
	\begin{align*}
	\sigma^2(t) &= \frac{\pt^2}{\pt s\pt s'}F_{t,t}(y(t),y(t)) - \left[\ptt F_t(y(t))\right]^2 + 2 y'(t)\left[\frac{\pt^2}{\pt s\pt z'} F_{t,t}(y(t),y(t)) \right.\\
	&\quad - \left.f_t(y(t))\ptt F_t(y(t))\middle]+ y'^{2}(t)\middle[f_{t,t}(y(t),y(t)) - f_t(y(t))^2\right].
	\end{align*}
\end{cor}
These results provide rates of convergence and theoretical justifications for the estimated rank dynamics. 

\section{Simulation}\label{sec:simu}

For the implementation of the dynamic model in Section \ref{sec:decom} and the summary measures in Section \ref{sec:sumstat}, two important auxiliary parameters $h_Y$ and $h_T$ are involved to obtain the kernel estimators for the rank trajectories $R_i(\cdot)$ and the two components, $C_1(t)$ and $C_2(t)$, of the rank derivatives.
In this section, we use simulations to evaluate the finite-sample performance of the bandwidth selection method in the Supplementary Material, and the kernel estimators for $C_1(t)$ and $C_2(t)$ in model \eqref{eq:rankDeriv}. 

Denote $\phi$ and $\Phi$ as the probability density function and cumulative distribution function of the standard Gaussian distribution. Suppose we observe trajectories $Y_i(t) = \sum_{k=1}^5 \xi_{ik} \psi_k(t)$ for subjects $i=1,\dots,n$ on a dense time grid $\{j/m: j=0,1,\dots,m\}\subset \cT=[0,1]$, where $\psi_1(t) = 6(t-0.5)^2 \mbf{1}_{\{t>0.5\}}$, $\psi_2(t) = 0.4 + (70/9)\phi((t-0.5)/0.09)$, $\psi_3(t) = 0.6\cos(8\pi t)$, $\psi_4(t) = \sin(2\pi t)+1$, $\psi_5(t) = 8 \phi((t-0.2)/0.05)$, $\xi_{i1} \sim \cN(1.4, 1.7^2)$, $\xi_{i2} \sim \cN(1, 0.6^2)$, $\xi_{i3} \sim \cN(0, 0.5^2)$, $\xi_{i4} \sim \cN(0.8, 0.4^2)$, and $\xi_{i5} \sim \cN(0.4, 0.2^2)$, independently across $i=1,\dots,n$.
Hence, the true values of $R_i(t)$, $C_{1i}(t)$ and $C_{2i}(t)$ are respectively
\begin{align*}
R_i(t) &= \Phi \left(\frac{\sum_{k=1}^5 (\xi_{ik} - \mu_k)\psi_k(t)}{\sqrt{\sum_{k=1}^5 \sigma_k^2\psi_k(t)^2}}\right),\\
C_{1i}(t) &= \left[ \frac{-\sum_{k=1}^5 \mu_k\psi'_k(t)}{\sqrt{\sum_{k=1}^5 \sigma_k^2\psi_k(t)^2}} - \frac{\left[\sum_{k=1}^5 (\xi_{ik} - \mu_k)\psi_k(t)\right] \left[\sum_{k=1}^5 \sigma_k^2\psi_k(t)\psi'_k(t)\right]} {\left[\sum_{k=1}^5 \sigma_k^2\psi_k(t)^2\right]^{3/2}}\right]\\
&\quad \cdot \phi\left(\frac{\sum_{k=1}^5 (\xi_{ik} - \mu_k)\psi_k(t)}{\sqrt{\sum_{k=1}^5 \sigma_k^2\psi_k(t)^2}} \right),\\
C_{2i}(t) &= \frac{\sum_{k=1}^5 \xi_{ik}\psi'_k(t)}{\sqrt{\sum_{k=1}^5 \sigma_k^2\psi_k(t)^2}} \cdot \phi\left(\frac{\sum_{k=1}^5 (\xi_{ik} - \mu_k)\psi_k(t)}{\sqrt{\sum_{k=1}^5 \sigma_k^2\psi_k(t)^2}} \right).
\end{align*}

To assess the performance of the cross-validation (CV) selected bandwidths $(h_Y^\cv, h_T^\cv)$, we compared the mean integrated squared errors (MISEs) of $\wt C_{1i}(t)$ and $\wt C_{2i}(t)$ obtained with the CV bandwidths as well as with  the optimal choice given by
\[(h^\opt_Y, h^\opt_T) = \argmin_{(h_Y, h_T)\in \mc{H}} \left[\mise (h_Y,h_T; \wt C_1) + \mise (h_Y,h_T; \wt C_2)\right],\] where $\mc H\in\R^2$ is the set of bandwidth pairs considered,
\begin{gather*}
\mise (h_Y,h_T; \wt C_1) = \fn \si1n \int_{h_{\max}}^{1-h_{\max}} \left[\wt C_{1i}(t) - C_{1i}(t)\right]^2 \diffop t,\\
\mise (h_Y,h_T; \wt C_2) = \fn \si1n \int_{h_{\max}}^{1-h_{\max}}  \left[\wt C_{2i}(t) - C_{2i}(t)\right]^2 \diffop t,
\end{gather*}
and $h_{\max}$ is the maximum value of $h_T$ considered. The impact of boundary effects is known to distort bandwidth selection and  is removed by cutting off $[0,h_{\max})$ and $(1-h_{\max},1]$ in the integration.

\begin{figure}[htbp]
	\centering
	\includegraphics[width=0.8\textwidth]{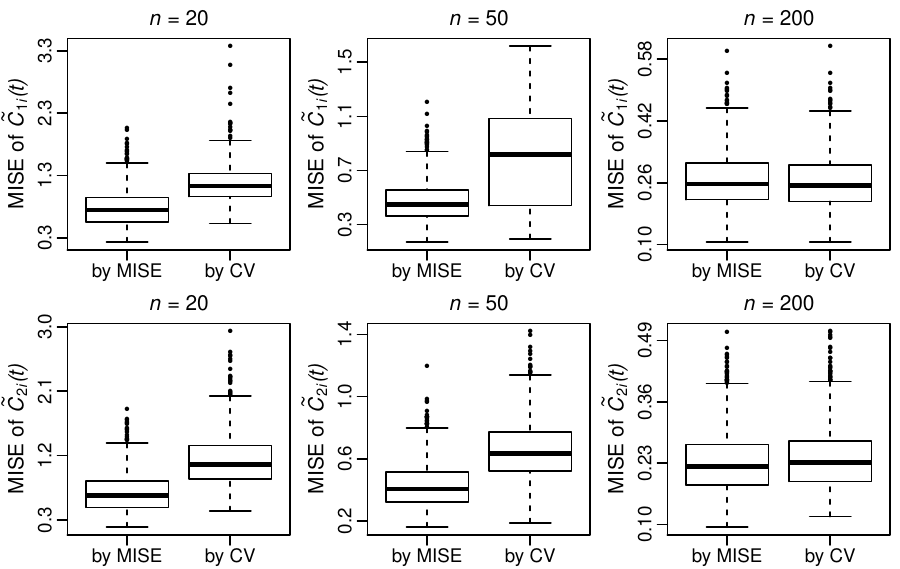}
	\caption{Boxplots of the MISEs of the estimators of the two rank derivative components $\wt C_{1i}(t)$ and $\wt C_{2i}(t)$ as per \eqref{eq:rankCompEst} corresponding to the optimal bandwidths chosen by MISE and CV in 1000 runs.}\label{fig:boxC1C2}
\end{figure}

In the simulations, we used $m=31$, $\mc H = \{(h_Y,h_T) = (2.4\times 0.6^u, 0.3\times 0.6^v): u,v=0,1,2,3\}$, and considered three different sample sizes $n=20$, 50 and 200. 
The kernels $K$ and $H$ used in Sections~\ref{sec:simu} and \ref{sec:app} are the pdf and cdf of standard normal distributions truncated on $[-4,4]$, respectively. Specifically,
\begin{align*}
K(x) &= \phi(x) \mathbf{1}_{[-1,1]}(x/4) / [\Phi(4) - \Phi(-4)], \quad\text{and}\\
H(x) &= [\Phi(x) - \Phi(-4)] \mathbf{1}_{[-1,1]}(x/4) / [\Phi(4) - \Phi(-4)] + \mathbf{1}_{(1,\infty)}(x/4),
\end{align*}
where $\phi$ and $\Phi$ are the pdf and cdf of standard normal distributions, respectively. We use these kernels as 
in practical implementations they yield smooth estimates $\widetilde{F}$, $\widetilde{D}_1$, and $\widetilde{D}_2$, while the pdf kernel $K$ has a compact support. 
Boxplots of the MISEs corresponding to the optimal bandwidths chosen by MISE and CV in each of the 1000 Monte Carlo runs for $n=20$, 50 and 200 are shown in Figure~\ref{fig:boxC1C2}.
The main message is that CV performs satisfactorily, as it tracks the optimal choice closely, especially for larger sample sizes  $n$. 

\begin{figure}[htbp]
	\centering
	\includegraphics[width=0.8\textwidth]{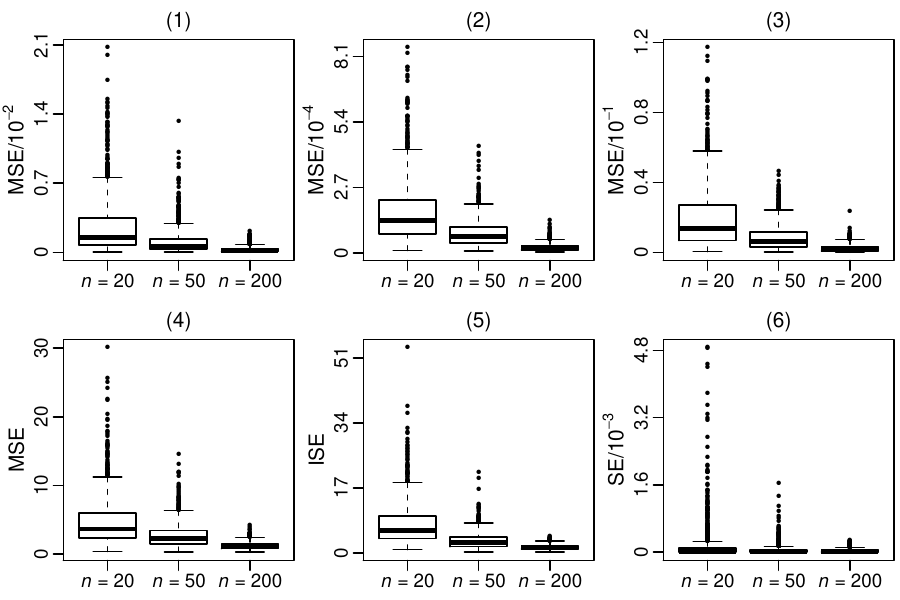}
	\caption{Boxplots of the MSEs, ISE or SE of various rank summary statistics 
		based on 1000 runs and CV bandwidths.  Panels  (1)--(6) display the results for the estimated values of the summary measures of rank processes $\rho_i$, $\nu_i$, $\zeta_i$, $\eta_i$, $\gamma(t)$, and $G$, as per \eqref{eq:intRank}--\eqref{eq:overallMixing}, respectively. 
	} \label{fig:boxTool}
\end{figure}

Boxplots of the MSEs, ISE or SE for the estimation of the rank summary measures \eqref{eq:intRank}--\eqref{eq:overallMixing} based on the kernel estimators $\wt R_i(t)$ and $\wt R'_i(t)$ obtained with the optimal bandwidths chosen by CV are shown in Figure~\ref{fig:boxTool}. Overall the proposed estimators are seen to converge fast to the  true values as $n$ increases.

\section{Applications}\label{sec:app}

We demonstrate our methods with  three functional datasets which are very different in nature. The first is the Z\"urich longitudinal growth data; the second is US  median house price data at the county level;  the third is based on the 2017 Major League Baseball (MLB) season, where our interest lies in offensive or hitting performances. We find that by transforming the original processes into rank processes we are able to find new and interesting characterizations for the individuals in each dataset. 

\subsection{Z\"urich Longitudinal Growth Data}

\begin{figure}[htbp]
	\centering
	\includegraphics[width=0.46\textwidth]{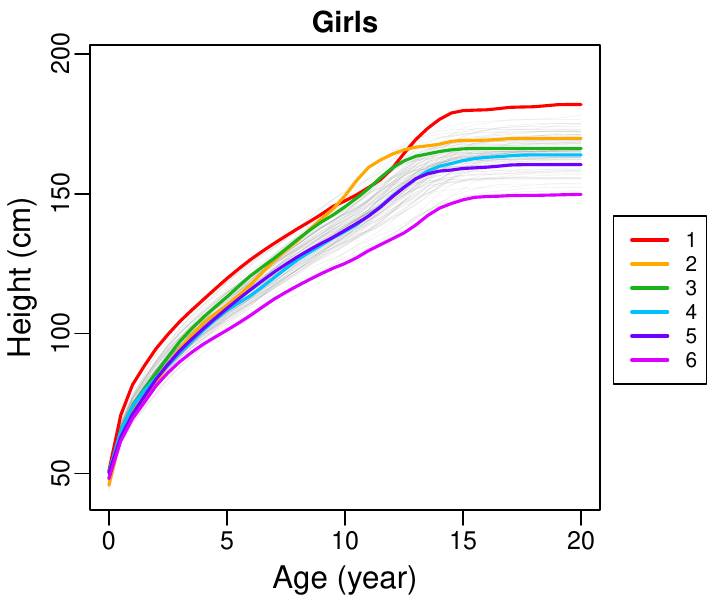}
	\includegraphics[width=0.46\textwidth]{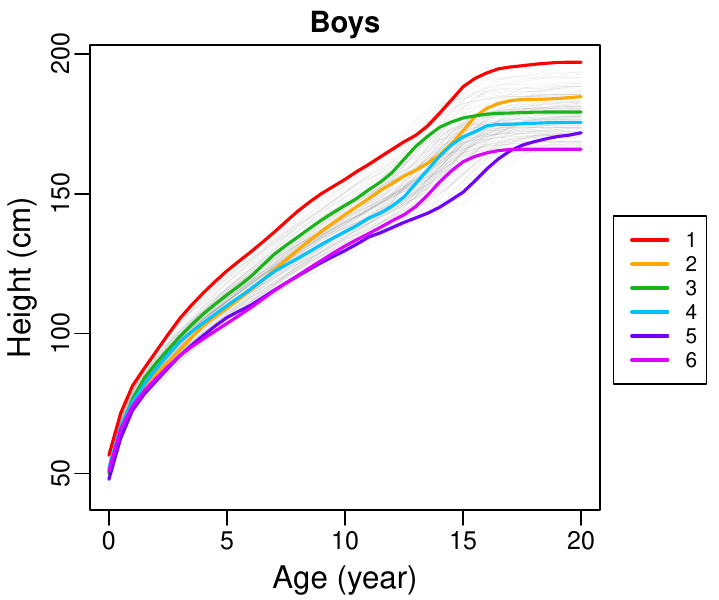}
	\caption[Z\"urich growth curves]{Pre-smoothed Z\"urich growth curves with six subjects highlighted for boys and girls.}
	\label{fig:zurData}
\end{figure}

The Z\"urich longitudinal growth data consist of dense longitudinal height measurements for 112 girls and 120 boys from birth to age 20 and the measurements are known to contain very little noise \citep{gass:90}.  It is helpful to compare the ranking for individuals; we highlight the same six girls and six boys throughout, with their  height trajectories shown in Figure~\ref{fig:zurData}.

\begin{figure}[htbp]
	\centering
	\includegraphics[width=0.46\textwidth]{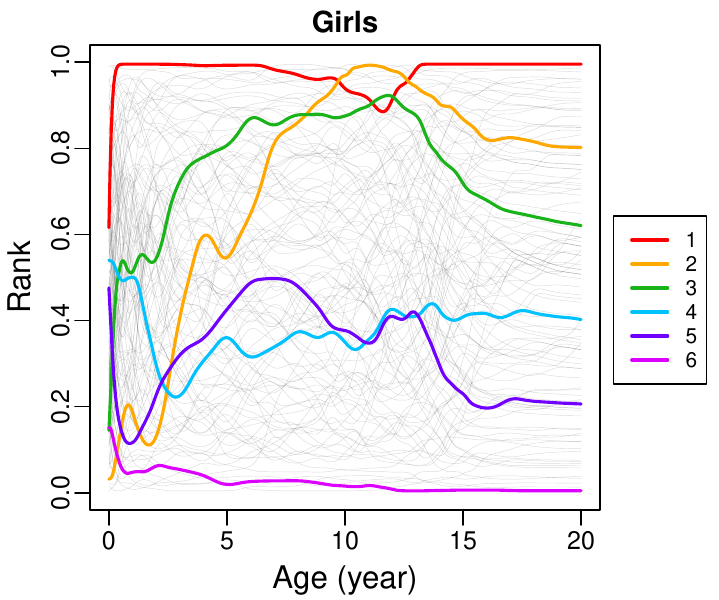}
	\includegraphics[width=0.46\textwidth]{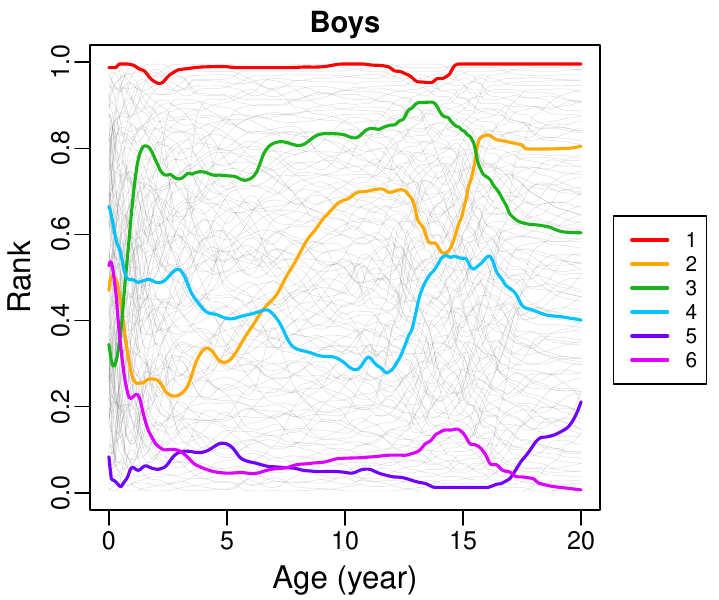}
	\caption[Smooth ranks for Z\"urich growth data]{Smoothly ranked Z\"urich growth trajectories.}
	\label{fig:zurSmooth}
\end{figure}

We find that the two ranking methods yield similar results, with the smooth rank functions resembling the empirical ranks. Visually, it is clear that taking a ranked perspective with functional data is appealing. For example, from Figure~\ref{fig:zurSmooth}, Girl 1 and Boy 1 are seen to be generally tall throughout, and Girl 2 and Boy 2 are seen to have volatile ranks as they age. Ranks are fairly stable from ages 5 until 10 and 12 for girls and boys, respectively; subsequently,  the ranks are more dynamic, with higher volatility.

\begin{figure}[htbp]
	\centering
	\includegraphics[width=0.46\textwidth]{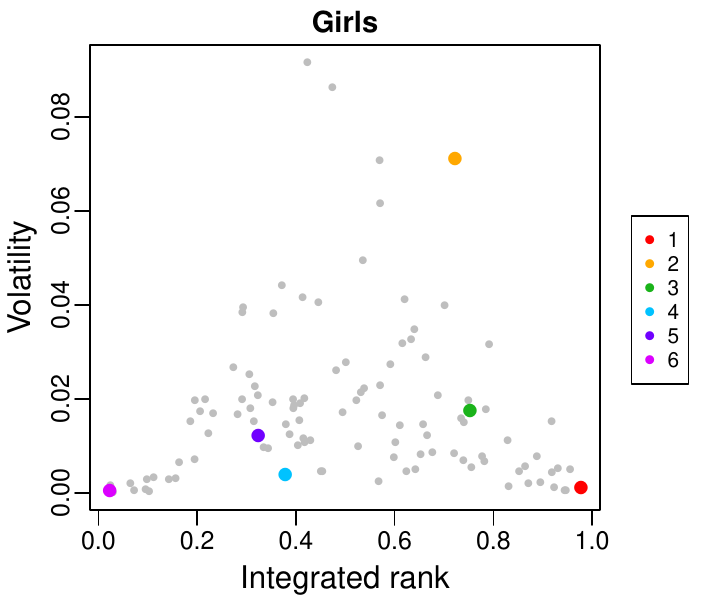}
	\includegraphics[width=0.46\textwidth]{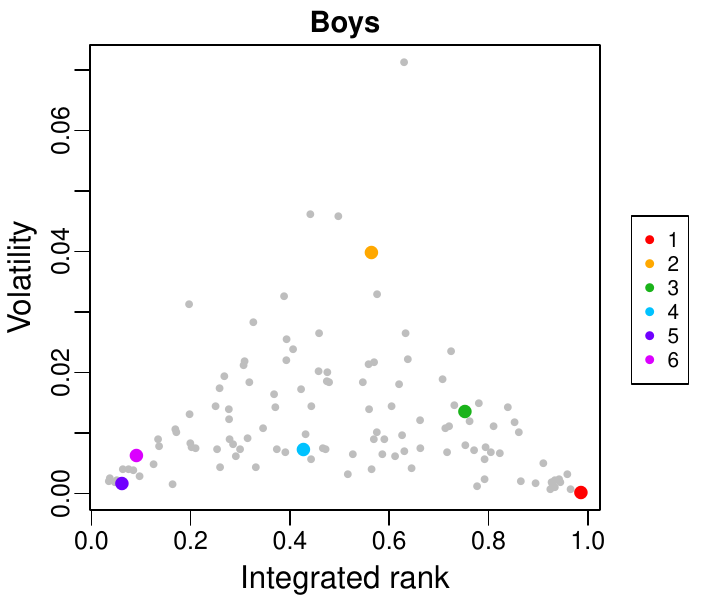}
	\caption[Rank volatility versus integrated rank in Z\"urich growth data]{Rank volatility $\nu_i$ as per \eqref{eq:intRankVar} versus integrated rank $\rho_i$ as per \eqref{eq:intRank} in the Z\"urich growth data, with the same six subjects highlighted as in Figures 3 and 4.}
	\label{fig:zurInstab}
\end{figure}

\begin{figure}[htbp]
	\centering
	\includegraphics[width=0.4\textwidth]{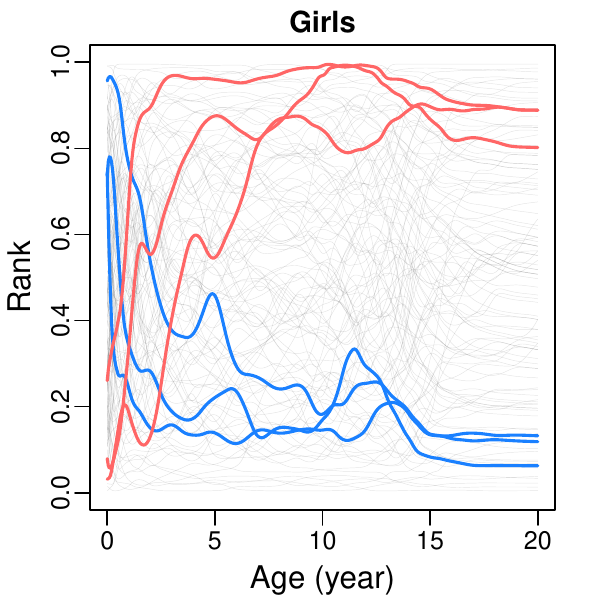}
	\includegraphics[width=0.4\textwidth]{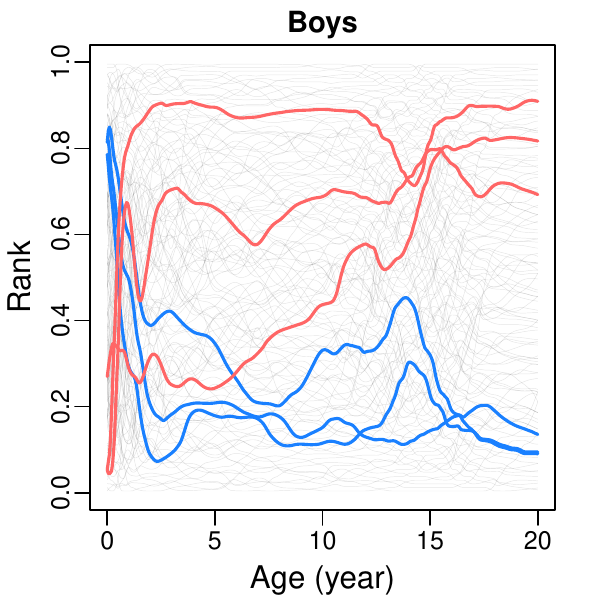}
	\caption[Ranked growth data for highest and lowest mixing coefficients]{Smoothly ranked Z\"urich growth data. Here we highlight the subjects with the highest (light red) and lowest (blue) subject-specific rank stability measures $\zeta_i$ as per \eqref{eq:mixing}.}
	\label{fig:zurMix}
\end{figure}

We also obtained the estimates of the rank summary statistics \eqref{eq:intRank}--\eqref{eq:overallMixing} for the Z\"urich longitudinal growth data, based on  the smooth ranks defined in \eqref{eq:smoothRank}. In Figure~\ref{fig:zurInstab} we see that Girl 1 and Boy 1 have very high ranks and that the ranks are almost constant throughout. On the other hand, we find that Girl 2 and Boy 2 have overall middle ranks that are quite volatile. The rank volatility plots are bell-shaped, as subjects with integrated ranks near 0 and 1 cannot have high volatility. On the other hand, subjects with moderate integrated ranks have less restricted volatility. We also highlight the subjects with the highest and lowest values of the subject-specific rank increases from start to end  $\zeta_i$ as in \eqref{eq:mixing} in Figure~\ref{fig:zurMix}, where $\zeta_i$  captures the overall ranking trend for a subject, i.e.,  subjects with large values of $\zeta_i$ have large increases or decreases in ranks from the beginning to the end of the time domain.

\begin{figure}[htbp]
	\centering
	\includegraphics[width=0.9\textwidth]{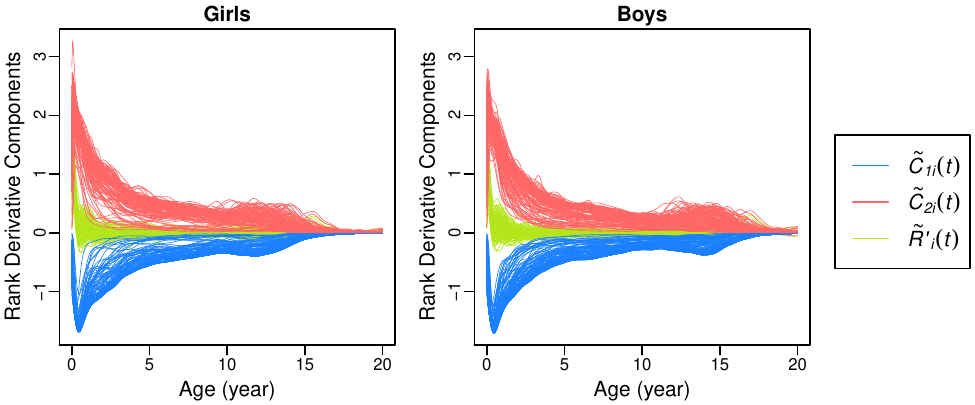}
	\caption[Rank derivative components for the Z\"urich growth data]{Estimated rank derivatives $\wt R_i'(t)$ and their two components $\wt C_{1i}(t)$ and $\wt C_{2i}(t)$ as per \eqref{eq:rankCompEst} for girls (left) and boys (right)  in the Z\"urich growth data.}
	\label{fig:zurComp}
\end{figure}

We also applied the rank decomposition  (\ref{eq:rankDeriv}) to the Z\"urich growth data. Figure~\ref{fig:zurComp} shows the rank derivative decomposition for all subjects in the study. The population trends quantified by the negative terms $C_1$ tend to lower an individual's rank as the population of children at large is growing, while individuals are also growing as reflected by the positive terms $C_2$. For the growth data, this decomposition indicates  that the population and individual components of the rank derivative are roughly equal in size. Indeed, the estimated contributions from the first component $\wt{\Lambda}_1$ for girls and boys are 0.487 and 0.486, with $\wt{\Lambda}_2 = 0.513$ and 0.514, respectively, for the second component.  We conclude that in human growth an individual's change in rank is the result of a fine balance of individual growth which is counterbalanced by population trends in growth when considering individual rank trajectories.   Rank volatility is seen to increase  during times of growth spurts, where the population tends to grow relatively fast while individuals may have accelerated or delayed growth, with resulting rank changes. 

\subsection{House Price Data}

\begin{figure}[htbp]
	\centering
	\includegraphics[width=0.6\textwidth]{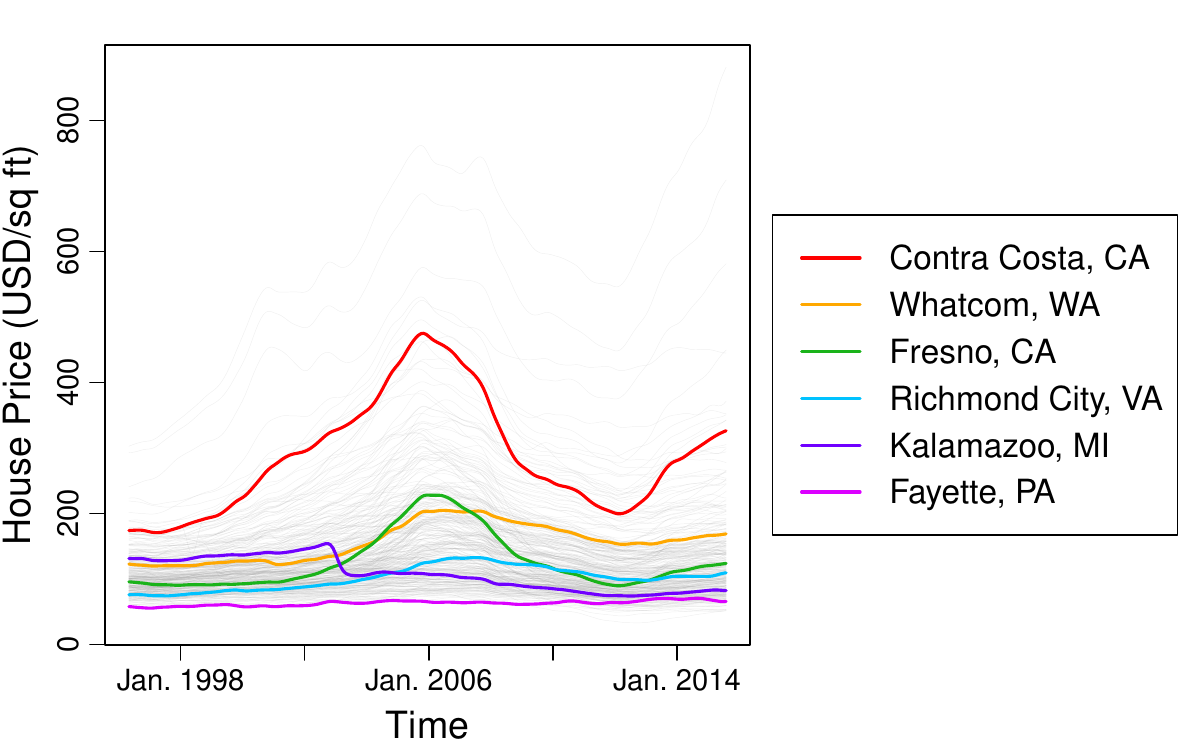}
	\caption[House price curves]{Pre-smoothed inflation-adjusted median house price curves from May 1996 to August 2015 for 306 US counties with six counties highlighted.}
	\label{fig:houData}
\end{figure}

\begin{figure}[htbp]
	\centering
	\includegraphics[width=0.6\textwidth]{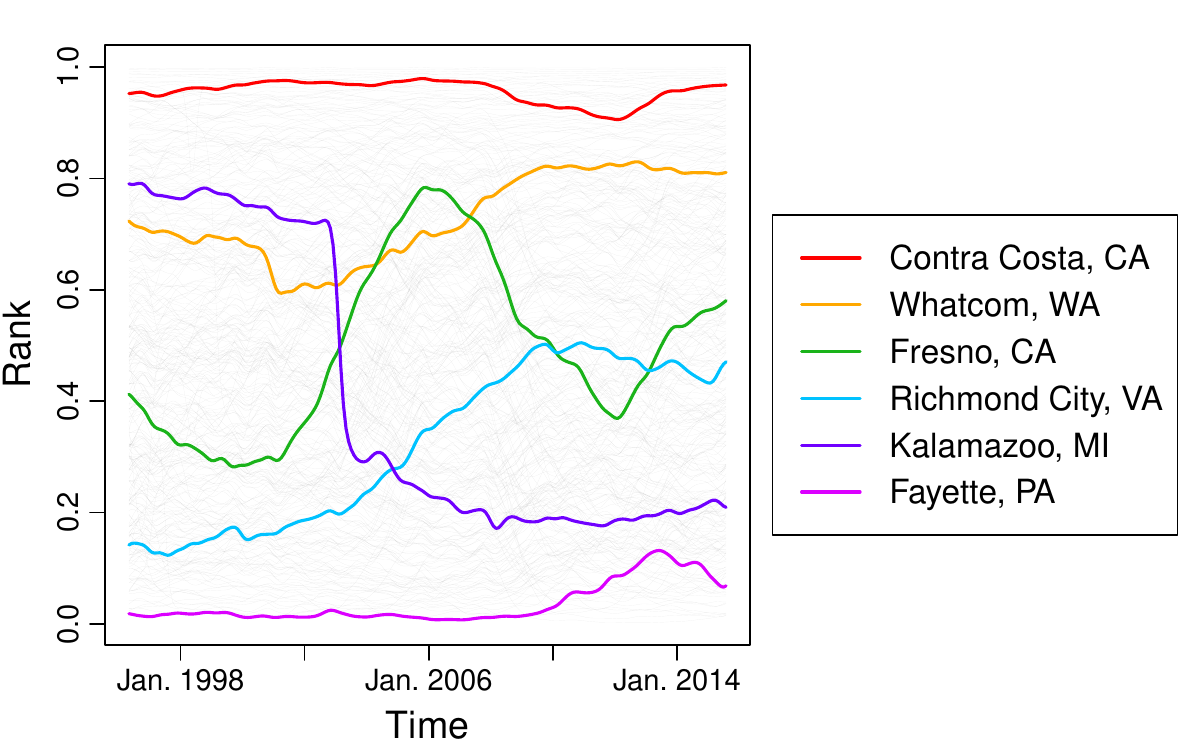}
	\caption[Smoothly embedded ranks for house price data]{Smoothly ranked house price trajectories.}
	\label{fig:houSmooth}
\end{figure}

House price data are available from Zillow. We consider here  monthly longitudinal median house prices  after inflation adjustment for house transactions in 306 counties  in the US from May 1996 to August 2015. To compare the ranking for individual markets, we highlight the same six counties  throughout, as in Figure~\ref{fig:houData}. Adopting  the smooth rank function version defined in \eqref{eq:smoothRank}, in Figure~\ref{fig:houSmooth} house prices in Contra Costa and Fayette   are seen to be generally high and low throughout, respectively, and those in Fresno  are seen to have significant rank variation.  We find that ranks were fairly stable before 2002 and became more dynamic afterward.

\begin{figure}[htbp]
	\centering
	\includegraphics[width=0.58\textwidth]{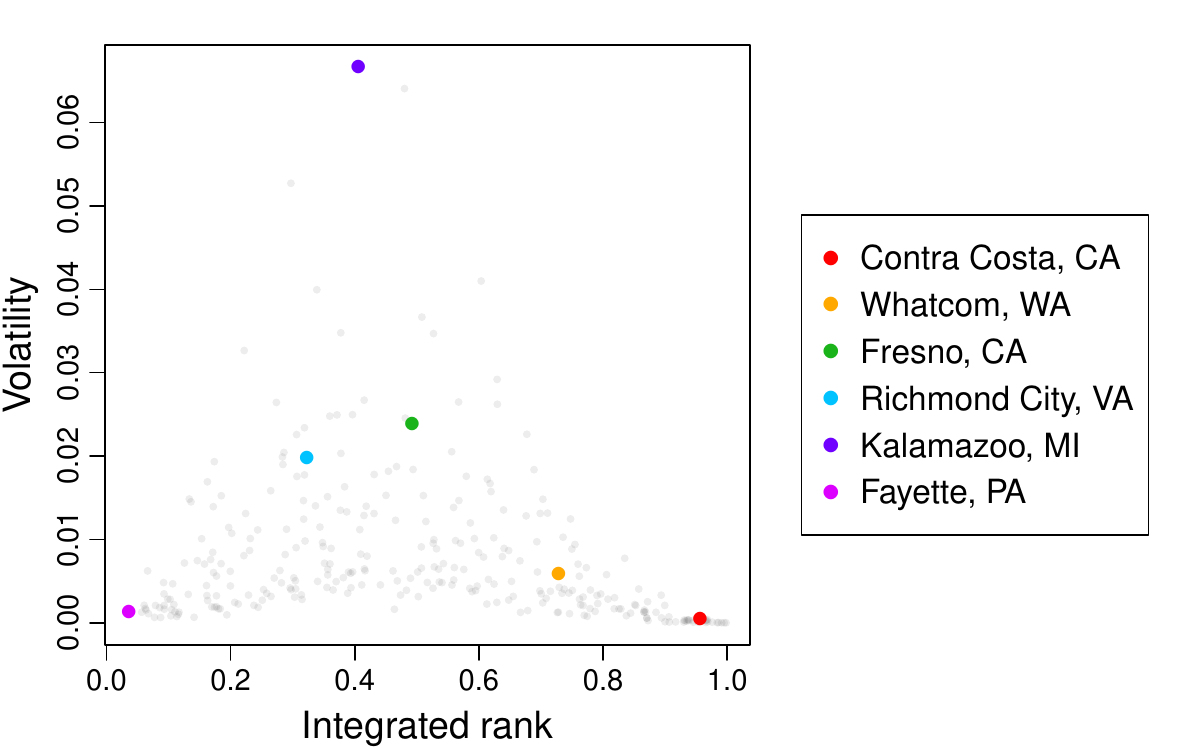}
	\caption[Volatility versus integrated rank in house price data]{Rank volatility $\nu_i$ as per \eqref{eq:intRankVar} versus integrated rank $\rho_i$ as per \eqref{eq:intRank} for the house price data, with the same six counties highlighted for clarity.}
	\label{fig:houInstab}
\end{figure}

\begin{figure}[htbp]
	\centering
	\includegraphics[width=0.58\textwidth]{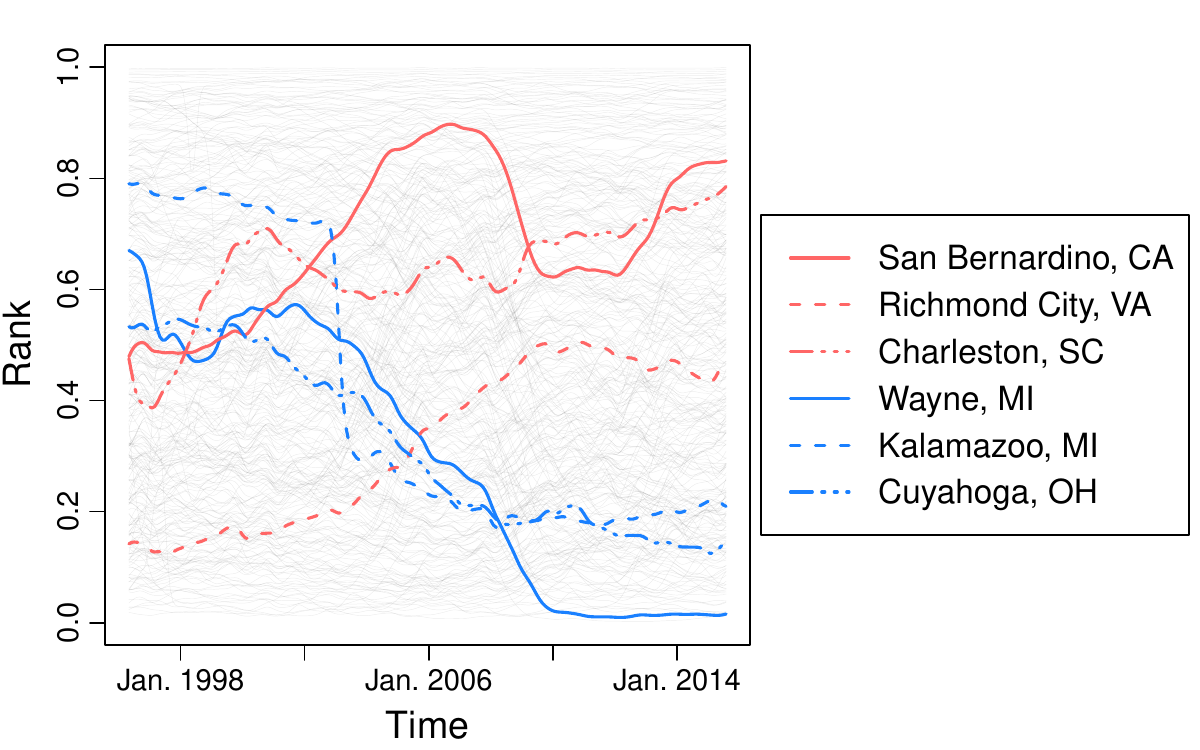}
	\caption[Ranked house price data for highest and lowest mixing coefficients]{Smoothly ranked house price data, highlighting  the counties  with the highest (light red) and lowest (blue) county-specific rank gains  $\zeta_i$ as per \eqref{eq:mixing}.}
	\label{fig:houMix}
\end{figure}

We also estimated the rank summary statistics for the house price data. In Figure~\ref{fig:houInstab} we see that Contra Costa county and Fayette county have very high and low ranks respectively and that their ranks were almost constant throughout the time period considered.  On the other hand, we find that Kalamazoo has moderate ranks that are very volatile. These findings are in agreement with Figure~\ref{fig:houSmooth} and the rank volatility plot has a similar shape to that in Figure~\ref{fig:zurInstab}, as expected. Highlighting the counties  with the highest and lowest gains in rank $\zeta_i$ as in \eqref{eq:mixing} in Figure~\ref{fig:houMix}, we find  that the magnitudes of difference in ranks between the beginning and the ending for the house price data are not as large as those for the Z\"urich growth curves.

\begin{figure}[htbp]
	\centering
	\includegraphics[width=0.8\textwidth]{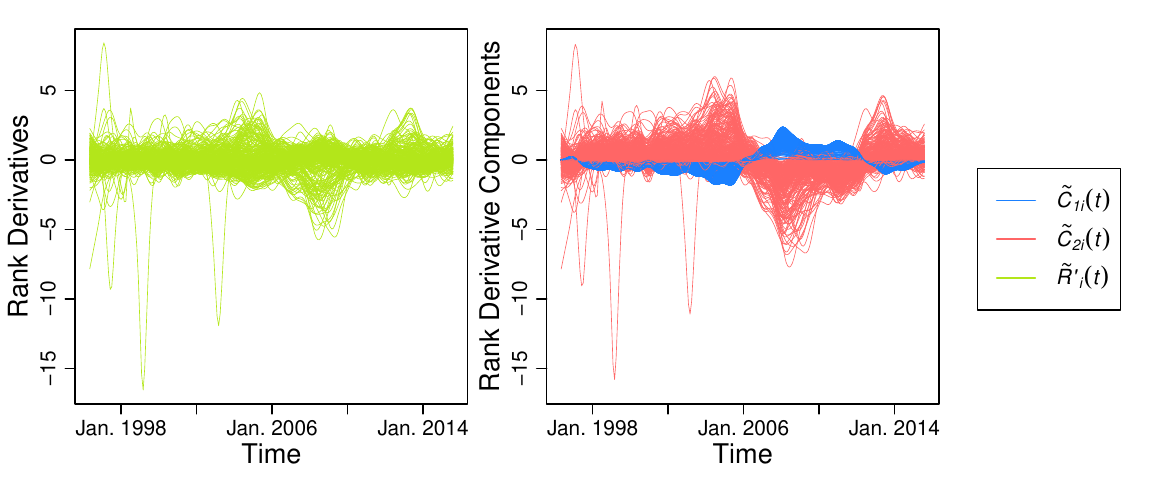}
	\caption[Rank derivative components for the house price data]{Estimated rank derivatives $\wt R_i'(t)$ and their two components $\wt C_{1i}(t)$ and $\wt C_{2i}(t)$ as per \eqref{eq:rankCompEst} for the house price data.}
	\label{fig:houComp}
\end{figure}

We also applied the dynamic rank decomposition (\ref{eq:rankDeriv}) to the house price curves. Figure~\ref{fig:houComp} shows the rank derivative decomposition for all counties in the study. The house price ranks were more volatile a few years before and after the 2008 financial crisis. The population components also reveal that  county median house prices were increasing in general before 2006, turned to drop from 2007, and then gradually recovered and increased  again since 2012. The individual component is seen to contribute more to the rank derivative than the population component. This is also reflected by the estimated contributions from the two components, $\wt{\Lambda}_1 = 0.458$ and $\wt{\Lambda}_2 = 0.542$, as per \eqref{eq:rankContriEst}. As shown in Figure~\ref{fig:houData}, a general trend cana be discerned from the house price trajectories: Prices  initially increased  until 2005, decreased from 2005 to 2012, and then increased again. The house price population dynamics points predominantly downwards until 2008, with individual markets exercising strong counterforces; this means a county  where price growth was sluggish fell back in rank; the opposite happened between 2008 and 2012 --- a county  where house prices were stable was gaining against the population and its rank increased.  

\subsection{Major League Baseball Offensive Data}

Another area where relative rank is important is in sports. Major League Baseball (MLB) teams routinely spend over \$100 million on player salaries every year. It is therefore of paramount interest to rank players in terms of ability so that teams can invest efficiently in individual players.  Although there are many factors which contribute to the overall value of a player, one of the most important is offensive performance, and accordingly we focus on ranking MLB players in terms of offense.


Baseball has recently become a game dominated by statistics \citep[see][and the movie \emph{Moneyball} for instance]{baum:13,silv:12}. As such, statisticians and sabermetricians look for simple yet informative measures for assessing player performance. By far, the most widely used statistic to quantify offensive performance is the batting average (BA), which is  the number of hits a player has divided by the number of attempts. While the batting average is simple to understand, it has several shortcomings; for example, late in the season, when the number of attempts or at-bats is high, the average will not easily reveal changes in performance. 

\begin{figure}[htbp]
	\centering
	\includegraphics[width=0.98\textwidth]{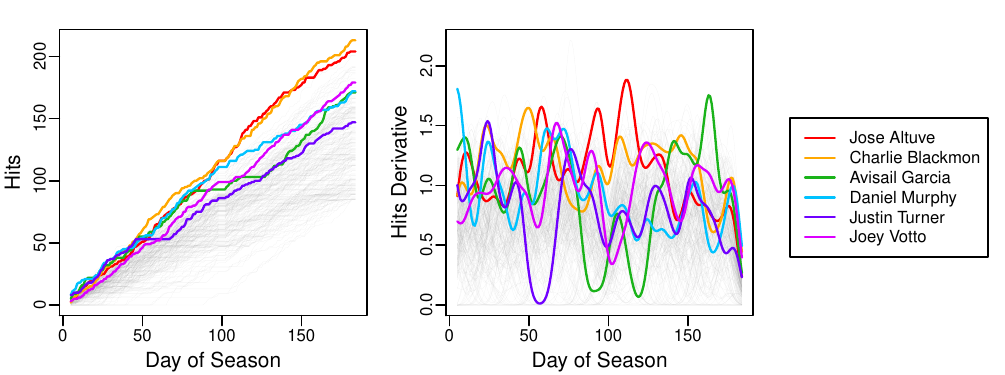}
	\caption[Cumulative hits and derivatives by day for the 2017 MLB data]{Cumulative hits (left) and hits derivatives (right) for each  day  in the 2017 Major League Baseball season for 237 players. Six curves are highlighted which correspond to the players with the highest batting averages.}
	\label{fig:basData}
\end{figure}

\begin{figure}[htbp]
	\centering
	\includegraphics[width=0.63\textwidth]{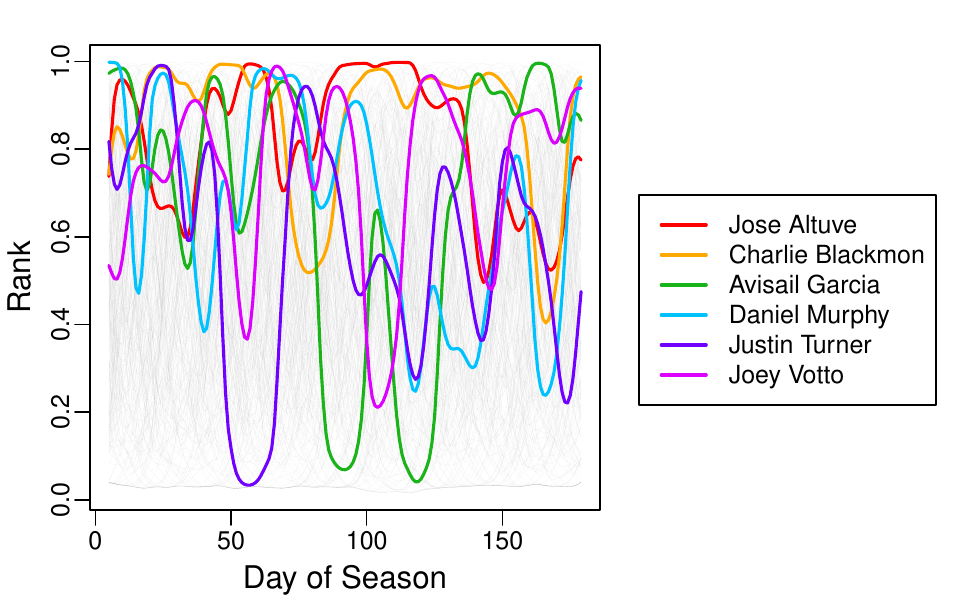}
	\caption[Rank transformed baseball data]{Rank transformed baseball data, with the same six players highlighted.}
	\label{fig:basRanks}
\end{figure}

In light of the drawbacks of using batting average as a response, we tracked  the number of hits a player accrued for  each day in the 2017 MLB season (\url{http://www.baseballmusings.com/}), and then took  the derivative of this trajectory, which we used as our functional response. This derivative can be viewed as  a local batting average, or the change in hits divided by the change in days. It is thus less  affected by long-term history because it is an instantaneous measure. This response therefore characterizes the \emph{heat} of a player, which is the level of their current  performance. 
The original hits trajectories and corresponding hits derivatives trajectories in Figure~\ref{fig:basData}, obtained by local polynomial smoothing, are our starting point for the rank analysis.  The objective  is to quantify the player's ranks and changes in ranks in this dataset, aiming to identify top players. 
We first transform the hit derivative trajectories into rank trajectories using the smooth representation in (\ref{eq:smoothRank}), visualized in Figure~\ref{fig:basRanks}, where the differences in rank for the six highlighted players are highlighted.  For example, this visualization makes it clear that Joey Votto improved drastically throughout the season, moving from a rank near 0.25 at the beginning of the season, to finishing with a rank of nearly 1.

\begin{figure}[htbp]
	\centering
	\includegraphics[width=0.6\textwidth]{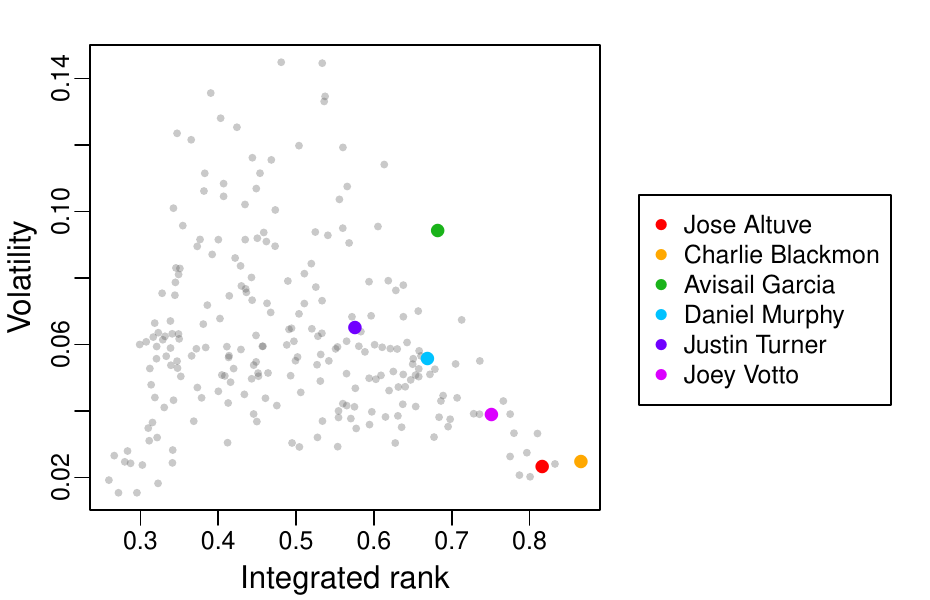}
	\caption[Rank volatility plot for baseball data]{Rank volatility $\nu_i$ as per \eqref{eq:intRankVar} versus integrated rank $\rho_i$ as per \eqref{eq:intRank} for the baseball data, with the same six players highlighted.}
	\label{fig:basInstab}
\end{figure}

We also applied the rank summary statistics, which prove to be  informative. The rank volatility versus integrated rank plot, shown in Figure~\ref{fig:basInstab}  has direct applications in assessing offensive performance from the 2017 MLB season. Naturally, all six of the highlighted players have relatively high ranks. In addition to average performance, we can see that two of the players, Jose Altuve and Charles Blackmon, had high integrated rank and low volatility, which are two features of the most valuable players. These players are consistently performing at a high level with respect to the rest of the sample. As shown in Figure~\ref{fig:basInstab}, the player with the highest integrated rank and fairly low volatility is Charlie Blackmon. 
Taking the viewpoint of a team deciding on which players to acquire, this plot also allows one to select players who have modest average ranks but have low volatility. Players of this type are desirable when looking for consistent backup players, for example. Finally,  the player-specific change in ranks $\zeta_i$ as per \eqref{eq:mixing} quantifies whether players are generally improving or deteriorating over the season.

\begin{figure}[htbp]
	\centering
	\includegraphics[width=0.8\textwidth]{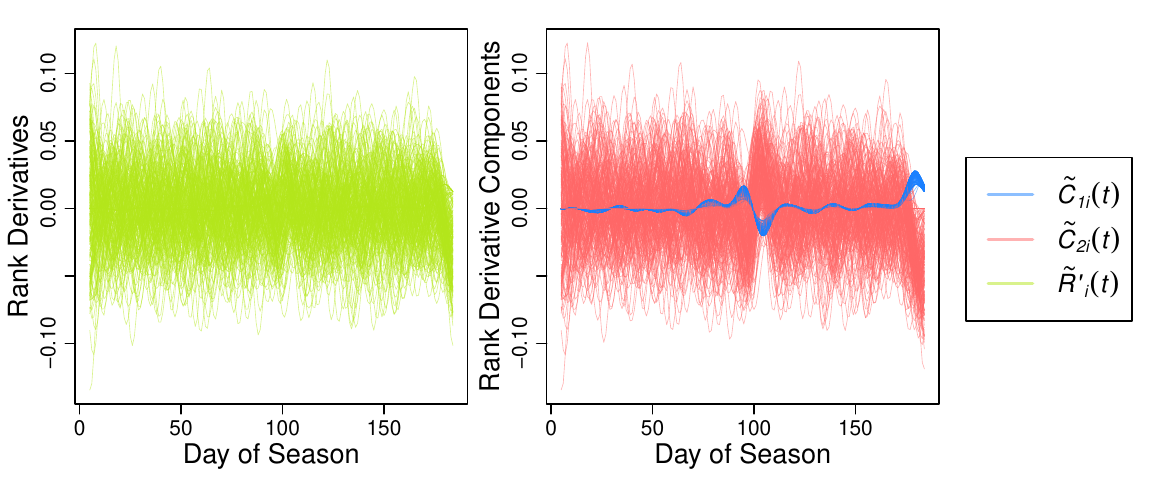}
	\caption[Rank derivative components for Major League Baseball data]{Estimated rank derivatives $\wt R_i'(t)$ and their two components $\wt C_{1i}(t)$ and $\wt C_{2i}(t)$ as per \eqref{eq:rankCompEst} for 2017 Major League Baseball data.}
	\label{fig:basComp}
\end{figure}

\begin{figure}[htbp]
	\centering
	\includegraphics[width=0.98\textwidth]{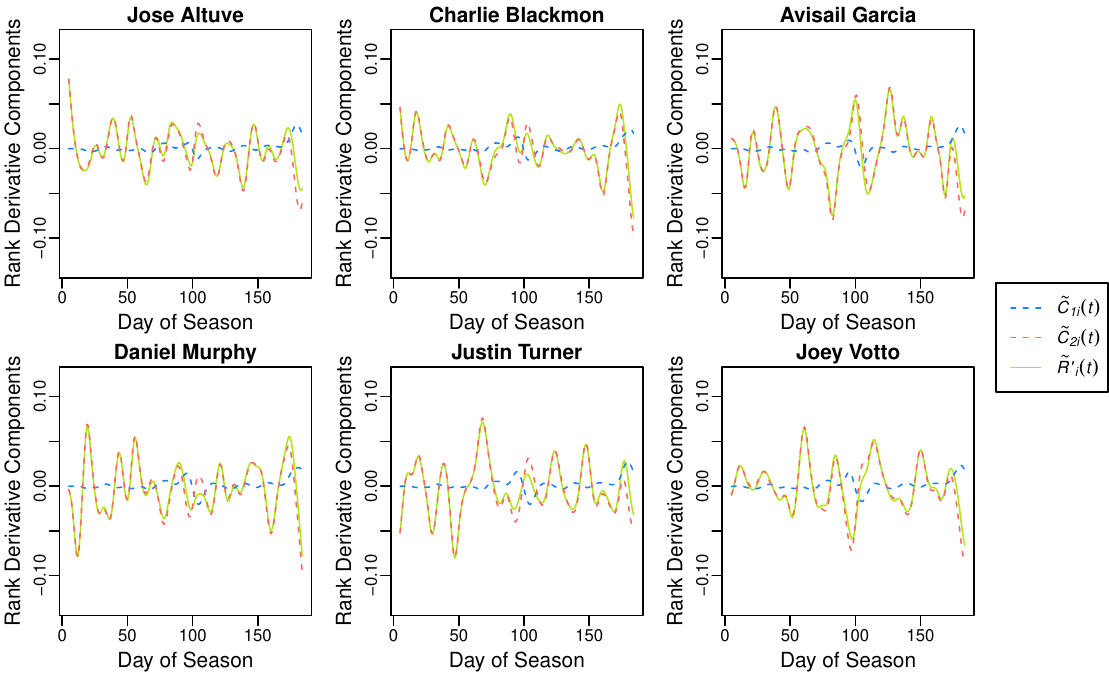}
	\caption[Rank derivative components for six top Major League players]{Estimated rank derivatives $\wt R_i'(t)$ and their two components $\wt C_{1i}(t)$ and $\wt C_{2i}(t)$ as per \eqref{eq:rankCompEst} for the six players with the highest batting averages of 2017.}
	\label{fig:basCompH}
\end{figure}

When fitting the rank derivative decomposition model (\ref{eq:rankDeriv}) to these baseball data,   
we find that the subject specific component $C_2(t)$ contributes much more than the population component $C_1(t)$. This is not surprising as  the population of hits derivative curves $Y_i(t)$, $i = 1, \dots, n$ does not have a very clear pattern. Thus rank is determined to a large extent by individual effort alone, with estimated contributions  $\wt{\Lambda}_1 = 0.165$ for the population component and $\wt\Lambda_2 = 0.835$ for the individual component. This  is visualized in  Figures~\ref{fig:basComp} and \ref{fig:basCompH}, where the second component is seen to dominate the first. In addition, an ascent followed by a descent period can be seen in the population component curves around Day 100. This is due to the ``All Star Break", which is a break for all the players except the All Stars. i.e.,  the best players from each team, who play in an exhibition game. Thus, the hits derivatives decrease toward zero for almost all players during the break and then recover after the games are resumed. Hence the population components first ascend and then descend accordingly. The ascending phase of the population component near the end of the season is due to the same reason, i.e.,  fewer games are available at that time.

Finally, the overall rank stability coefficient $G$ in \eqref{eq:overallMixing}, which is an overall scaled measure of how variable the rank trajectories are, can be used to compare all three functional data set  that we have considered, i.e.,  the Z\"urich growth data, the housing price data and the baseball player data.  The estimates of $G$ based on the smooth rank estimation are shown in Table \ref{tab:G}. The baseball players' rank curves have the lowest stability, with the most volatility of ranks and  a much higher degree of crossing trajectories. Moreover, the  rank trajectories  are not much influenced by population trends. In the Z\"urich growth and house price data, we observe much higher degrees of stability, with the highest level of rank stability and associated lowest rank volatility for the growth data. Especially for the growth data, crossings of rank trajectories are not common.  Rank trajectories for the housing data and even more so for the growth data are driven to a large extent by population trends, where population distributions uniformly move to higher levels for the growth data with increasing age, while they have increasing and decreasing phases for the house price data. Notably, for the growth data, the trajectory dynamics are driven in equal parts by population trends and individual growth patterns,  while for the housing price data population trends play a slightly smaller role. 

\begin{table}[htbp]
	\centering
	\caption{Estimates of $G$ based on the smooth rank estimation for all the three datasets}{
		\begin{tabular}{cccc}
			\toprule
			\multicolumn{2}{c}{Z\"urich growth} & \multirow{2}{*}{House price} & \multirow{2}{*}{Baseball}\\
			Girls & Boys & &\\
			\hline
			0.9866 & 0.9883 & 0.4500 & $3.409\times 10^{-21}$\\
			\bottomrule
	\end{tabular}}\label{tab:G}
\end{table}

\section{Discussion}\label{sec:disc}

Cross-sectional ranking of functional data is a powerful tool for exploratory functional data analysis. To the best of our knowledge, the proposed perspectives in this paper are new to the field of functional data analysis and allow for quantification of the rank dynamics of a stochastic process. These methods are simple to understand and straightforward to implement.  The decomposition of rank dynamics into population and individual components allows to better understand the forces that shape observed rank trajectories, and the summary measures of rank volatility, rank stability  and rank gain are useful. 

For the estimation of the two components $D_1(y,t)$ and $D_2(y,t)$ in \eqref{eq:ptDeriv}, we could alternatively use local quadratic regression. This would be asymptotically equivalent to the kernel estimator in \eqref{eq:estD} under regularity assumptions on the smoothness of weight functions and the shape of kernels \citep{mull:87:4}. However, the kernel method we employ here  has an explicit form which facilitates theoretical derivations, and makes  implementation straightforward, while the local quadratic regression involves the inverse of a matrix of dimension at least $5\times 5$.  This provides strong motivation for the proposed method.

Our estimation methods and theory are geared towards  densely observed functional data. One possible approach for the case of sparsely observed functional data is to divide the time domain into bins in a preprocessing step, followed by  estimating the cross-sectional distribution at time $t$ by using local Fr\'echet regression \citep{pete:19} based on the preliminary  distributions observed  at the midpoints of the bins -- these  are the empirical distributions derived from the  observations falling into each bin. The two components can then be obtained, e.g., by taking difference quotients of the cross-sectional distribution estimates. To work out the details and full theoretical justification of such a method will be a future research project.


\section*{Acknowledgements}
Funding: This work was supported by NSF Grant DMS-1712864.

\section*{Appendices}
\appendix
\renewcommand{\theequation}{S.\arabic{equation}}

\section{Bandwidth Selection for the Kernel Estimator}
It is important to provide a data-driven approach for bandwidth selection for the kernel estimator in \eqref{eq:estF}. For a complete discussion on optimal bandwidth selection for nonparametric conditional distribution and quantile functions, see \citet{li:13}. A simple objective function used in this paper is
\be\nn
CV(h_Y, h_T) \coloneqq \sum_{\begin{subarray}{c} 1\le i\le n,\ 1\le j\le m_i,\\ t_{ij}\in (h_{\max}, 1-h_{\max})\end{subarray}} \int_{-\infty}^\infty \left[1(\Yij \leq y) - \wt{F}_{\tij,-(i,j)}(y)\right]^2 dy,
\ee
where $h_{\max}$ is the maximum value considered for $h_T$, and $\wt{F}_{\tij,-(i,j)}(y)$ is the leave-one-out kernel estimator.

Alternative methods for selecting bandwidths include independently choosing the optimal bandwidths in the $t$ and $y$ directions, and also using cross-validation schemes for bandwidth selection in the nonparametric cross-sectional distribution estimation. 
To accelerate the cross-validation process, $k$-fold cross validation can be used instead when the sample size $n$ is relatively large. 
One can also perform the cross-validation on a random subset  with indices $\subset \{(i,j):1\le i\le n, 1\le j\le m_i\}$ rather than the entire sample. 
Another method is changing the sampling unit in the cross validation from single pairs $(t_{ij}, Y_{ij})$ to trajectories/subjects, i.e., 
performing one-curve-left-out cross-validation or assigning subsets of pairs $\{(t_{ij},Y_{ij}):\ j=1,\ldots, m_i\}$ to training or test sets in $k$-fold cross-validation.

\section{Details on Theoretical Results}
To derive the asymptotic normality for $\wt F_t(y)$, $\wt D_1(y,t)$ and $\wt D_2(y,t)$, we first need to calculate the means, variances of and (some) covariances between $Q_{li}(y,t)$, for $l=1,\ldots,5$.
For completeness, we include auxiliary Lemma \ref{lem:int01} and \ref{lem:kerXY}, which are well-known. Proof can be found in, e.g., \citet{rose:56}, \citet{rous:69}, \citet{silv:86}, \citet{sama:89}, \citet{stok:93}, \citet{jone:94}, and \citet{cai:99}.

\begin{lem}\label{lem:int01}
	Assume \ref{ass:a1} and \ref{ass:a2}. Arbitrarily fix $z\in(0,1)$. Suppose $g$ is a three times continuously differentiable function on $[0,1]$. For $h \le \min\{z,1-z\}$, as $h\ra 0$,
	\begin{gather*}
	\i01 h^{-1} g(x) K\left(\frac{z-x}{h}\right) dx = g(z) + \half h^2 \sigma^2(K) g''(z) + o(h^2),\\
	\i01 h^{-2} g(x) K'\left(\frac{z-x}{h}\right) dx = g'(z) + \half h^2\sigma^2(K)g^{(3)}(z) + o(h^2).
	\end{gather*}
	Now suppose $g(x,y)$ is a four times continuously differentiable function on $[0,1]^2$. Denote
	\[\frac{\pt^{k+l}}{\pt x^k\pt y^l}g(z,z') = \left.\frac{\pt^{k+l} g(x,y)}{\pt x^k\pt y^l}\right|_{\begin{subarray}{l} x=z\\ y=z' \end{subarray}}.\]
	For $h \le \min\{z,1-z\}$, as $h\ra 0$,
	\begin{align*}
	\i01\i01 h^{-2} g(x,y) K\left(\frac{z-x}{h}\right) K\left(\frac{z-y}{h}\right) dxdy
	&= g(z,z) +O(h^2),\\
	\i01\i01 h^{-3} g(x,y) K\left(\frac{z-x}{h}\right) K'\left(\frac{z-y}{h}\right) dxdy
	&= \frac{\pt}{\pt y}g(z,z) +O(h^2),\\
	\i01\i01 h^{-4} g(x,y) K'\left(\frac{z-x}{h}\right) K'\left(\frac{z-y}{h}\right) dxdy
	&= \frac{\pt^2}{\pt x\pt y} g(z,z) +O(h^2).
	\end{align*}
\end{lem}
\begin{lem}\label{lem:kerXY}
	Let $(X,Y)$ be a 2-dimensional random vector. Assume that the joint density and cdf of $(X,Y)$, $f_{XY}$ and $F_{XY}$, are twice and three times continuously differentiable, respectively. Under assumption \ref{ass:a1}, for any $h>0$, as $h\ra 0$,
	\begin{align*}
	&\E \left[h^{-2} K\left(\frac{z-X}{h}\right) K\left(\frac{z-Y}{h}\right)\right]\\
	&= f_{XY}(z,z) + \half h^2\sigma^2(K) \left[ \frac{\pt^2}{\pt x^2} f_{XY}(z,z) + \frac{\pt^2}{\pt y^2} f_{XY}(z,z)\right] + o(h^2),\\
	&\E \left[h^{-1} H\left(\frac{z-X}{h}\right) K\left(\frac{z-Y}{h}\right)\right]\\
	&= \frac{\pt}{\pt y}F_{XY}(z, z) + \half h^2 \left[\sigma^2(K) \frac{\pt^3}{\pt y^3} F_{XY}(z, z) + \sigma^2(H') \frac{\pt}{\pt x} f_{XY}(z, z)\right] + o(h^2),\\
	&\E \left[H\left(\frac{z-X}{h}\right) H\left(\frac{z-Y}{h}\right)\right]\\
	&= F_{XY}(z, z) + \half h^2 \sigma^2(H') \left[\frac{\pt^2}{\pt x^2} F_{XY}(z,z) + \frac{\pt^2}{\pt y^2} F_{XY}(z, z)\right] + o(h^2).
	\end{align*}
\end{lem}
Define
\begin{align*}
Q_{1i}(y,t) &=
\i01 h_T^{-1} H\left\{\frac{y-\Ys}{h_Y}\right\} K\left(\frac{t-s}{h_T}\right)d\theta(s),\\
Q_{2i}(y,t) &= \i01 h_T^{-1} K\left(\frac{t-s}{h_T}\right)d\theta(s),\\
Q_{3i}(y,t) &= \i01 h_T^{-2}H\left\{\frac{y-\Ys}{h_Y}\right\} K'\left(\frac{t-s}{h_T}\right)d\theta(s), \\
Q_{4i}(y,t) &= \i01 h_T^{-2} K'\left(\frac{t-s}{h_T}\right)d\theta(s),\\ 
Q_{5i}(y,t) &= \i01 h_Y^{-1}h_T^{-1} K\left\{\frac{y-\Ys}{h_Y}\right\} K\left(\frac{t-s}{h_T}\right)d\theta(s), 
\end{align*}
and for $l=1,\ldots,5$,
\be\nn
Q_l(y,t) = \fn\si1n  Q_{li}(y,t).
\ee
\begin{lem}\label{lem:BiasVar}
	Assume \ref{ass:a1}--\ref{ass:a5}. Furthermore, assume that the cross-sectional density $f_t(y)$ and cdf $F_t(y)$ at any time $t$ are twice and three times continuously differentiable, respectively, and that the joint density $f_{s,s'}(z,z')$ and cdf $F_{s,s'}(z,z')$ at times $s$ and $s'$ are continuous and twice continuously differentiable, respectively. Arbitrarily fix $t\in(0,1)$ and $y\in\R$. For $h_Y>0$ and $h_T\in (0,\min\{t,1-t\}]$, as $h_Y,h_T\ra 0$: For $q_l = \E Q_{li}(y,t)$, $l=1,\dots,5$,
	\begin{align*}
	q_1 &= F_t(y)\theta'(t) + \half h_Y^2\sigma^2(H') \pty f_t(y)\theta'(t) + \half h_T^2 \sigma^2(K) \p2t [F_t(y)\theta'(t)] \\
	&\quad + o(h_Y^2) + o(h_T^2),\\
	q_2 &= Q_{2i}(y,t) = \theta'(t) + \half h_T^2\sigma^2(K) \theta^{(3)}(t) + o(h_T^2),\\
	q_3 &= \ptt [F_t(y)\theta'(t)] + \half h_Y^2\sigma^2(H') \ptt \left[\pty f_t(y) \theta'(t)\right] + \half h_T^2\sigma^2(K) \frac{\pt^3}{\pt t^3} [F_t(y)\theta'(t)] \\
	&\quad + o(h_Y^2) + o(h_T^2),\\
	q_4 &= Q_{4i}(y,t) = \theta''(t) + \half h_T^2\sigma^2(K) \theta^{(4)}(t) + o(h_T^2),\\ 
	q_5 &= f_t(y)\theta'(t) +\half h_Y^2 \sigma^2(K) \pysq f_t(y) \theta'(t) +\half h_T^2 \sigma^2(K) \p2t [f_t(y)\theta'(t)] \\
	&\quad + o(h_Y^2) + o(h_T^2). 
	\end{align*}
	For $v_l = \var[Q_{li}(y,t)]$, $l=1,3,5$,
	\begin{align*}
	v_1 &= [F_{t,t}(y,y) - F_t(y)^2 ]\theta'(t)^2 +o(1),\\
	v_3 &= \frac{\pt^2}{\pt s\pt s'}F_{t,t}(y,y)\theta'(t)^2 + \left[\frac{\pt}{\pt s}F_{t,t}(y,y) + \frac{\pt}{\pt s'}F_{t,t}(y,y)\right]\theta'(t)\theta''(t) + F_{t,t}(y,y)\theta''(t)^2\\
	&\quad  - \left[\ptt [F_t(y)\theta'(t)]\right]^2+o(1),\\
	v_5 &= [f_{t,t}(y,y) - f_t(y)^2]\theta'(t)^2 +o(1).
	\end{align*}
	For $v_{kl} = \cov[Q_{ki}(y,t), Q_{li}(y,t)]$, $k,l=1,3,5$,
	\begin{align*}
	v_{13} &= \frac{\pt}{\pt s'} F_{t,t}(y,y)\theta'(t)^2 + F_{t,t}(y,y)\theta'(t)\theta''(t) - F_t(y)\theta'(t) \ptt [F_t(y)\theta'(t)]+o(1),\\
	v_{15} &= \frac{\pt}{\pt z'} F_{t,t}(y,y)\theta'(t)^2 - F_t(y)f_t(y)\theta'(t)^2 +o(1),\\
	v_{35} &= \frac{\pt^2}{\pt s\pt z'} F_{t,t}(y,y)\theta'(t)^2 + \frac{\pt}{\pt z'} F_{t,t}(y,y)\theta'(t)\theta''(t) - f_t(y)\theta'(t) \ptt [F_t(y)\theta'(t)] +o(1).
	\end{align*}
\end{lem}
Lemma \ref{lem:BiasVar} follows directly from Lemmas \ref{lem:int01} and \ref{lem:kerXY}, so we omit the proof here.
\begin{lem}\label{lem:riem}
	Assume \ref{ass:a0}--\ref{ass:a3}. Arbitrarily fix $t\in(0,1)$ and $y\in\R$. As $m_i\ra \infty$ and $h_Y,h_T\ra 0$,
	\begin{align*}
	\wt Q_{1i}(y,t) - Q_{1i}(y,t) &= O_p\left(m_i\inv h_T^{-2}\right) +  O_p\left(m_i\inv h_Y\inv h_T\inv \right),\\
	\wt Q_{2i}(y,t) - Q_{2i}(y,t) &= O\left(m_i\inv h_T^{-2} \right),\\
	\wt Q_{3i}(y,t) - Q_{3i}(y,t) &= O_p\left(m_i\inv h_T^{-3} \right) +  O_p\left(m_i\inv h_Y\inv h_T^{-2} \right),\\
	\wt Q_{4i}(y,t) - Q_{4i}(y,t) &= O\left(m_i\inv h_T^{-3} \right),\\
	\wt Q_{5i}(y,t) - Q_{5i}(y,t) &= O_p\left(m_i\inv h_Y^{-2} h_T\inv \right) + O_p\left(m_i\inv  h_Y\inv h_T^{-2}\right),
	\end{align*}
	where all the $O_p$ and $O$ terms are uniform over all $i=1,\ldots,n$.
\end{lem}
\bpf
We prove the convergence rate of $\wt Q_{1i}(y,t) - Q_{1i} (y,t)$ as the other proofs are analogous. Define
\[ g_1(s) = h_T\inv H\left\{\frac{y-\Ys}{h_Y}\right\} K\left( \frac{t - s}{h_T}\right). \]
The derivative of $g_1$ is computed as
\begin{align*}
g'_1(s) &= - h_Y\inv h_T\inv H'\left\{\frac{y-\Ys}{h_Y}\right\} Y'_i(s) K\left( \frac{t - s}{h_T}\right) - h_T^{-2} H \left\{\frac{y-\Ys}{h_Y}\right\} K'\left( \frac{t - s}{h_T}\right)\\
&= O_p(h_Y\inv h_T\inv) + O_p(h_T^{-2}),
\end{align*}
where all the $O_p$ terms are uniform over $s\in[0,1]$ and $i=1,\ldots,n$.  
Hence, with $s_{ij} \in \left(\theta\inv\{(j-1)/m_i\}, \theta\inv(j/ m_i)\right)$ such that $\fm g_1(s_{ij}) =  \int_{(j-1)/m_i}^{j/m_i} g_1(s)d\theta(s)$ and $s^*_{ij}$ lying between $\tij$ and $s_{ij}$ such that $g_1(\tij) - g_1(s_{ij}) = g_1'(s^*_{ij}) (\tij - s_{ij})$,
\begin{align*}
\left|\wt Q_{1i}(y,t) - Q_{1i} (y,t)\right|
&= \left|\fm \sj1m g_1(\tij) - \i01 g_1(s)d\theta(s)\right|\\
&\le \frac 1 {m_i} \sj1m |g_1'(s^*_{ij})| \left[\theta\inv\left(\frac{j}{m_i}\right)- \theta\inv\left(\frac {j-1} {m_i}\right)\right]\\
&= O_p(m_i\inv h_Y\inv h_T\inv) + O_p(m_i\inv h_T^{-2}),
\end{align*}
since
\[\sj1m \left[\theta\inv\left(\frac{j}{m_i}\right)- \theta\inv\left(\frac {j-1} {m_i}\right)\right]
\le  \sj1m \fm \max_{t\in[0,1]}\theta^{-1'}(t) \le a_1\inv.\]
\epf
\bco
\bpf[Proof of Proposition 1]
By Lemma \ref{lem:BiasVar}, in order to balance the orders of the bias and variance of $Q_1(y,t)$, the optimal bandwidths are of the order
$h_Y\sim n^{-1/4}$ and $h_T \sim n^{-1/4}$.
Again by Lemma \ref{lem:BiasVar}, 
\begin{gather*}
Q_1(y,t) = F_t(t)\theta'(t) + O_p(n^{-1/2}),\quad \oline Q_1(y,t) - Q_1(y,t) = O_p\left(\sqrt{\fn \sum_i m_i^{-2}}\right),\\
Q_2(y,t)= q_2 = \theta'(t) + O(n^{-1/2}),\quad \oline Q_2(y,t) - Q_2(y,t) = O\left(\sqrt{\fn \sum_i m_i^{-2}}\right).
\end{gather*}
By Lemma \ref{lem:riem}, as $n,m_i\ra \infty$,
\begin{align*}
\sqrt n \left[\wt F_t(y) - \frac{Q_1(y,t)}{q_2}\right]
&= \sqrt n \left[\frac{\oline Q_1(y,t)}{\oline Q_2(y,t)} - \frac{Q_1(y,t)}{\oline Q_2(y,t)} + \frac{Q_1(y,t)}{\oline Q_2(y,t)} - \frac{Q_1(y,t)}{q_2} \right]\\
&= \sqrt n \left[\frac{\oline Q_1(y,t) - Q_1(y,t)}{\oline Q_2(y,t)} -  \frac{Q_1(y,t)\left[\oline Q_2(y,t) -  q_2\right]}{\oline Q_2(y,t)q_2}\right]\\
&= O_p\left(\sqrt{\sum_i m_i^{-2}}\right).
\end{align*}
Thus, as $n,m_i\ra\infty$,
\[\sqrt n \left[\wt F_t(y) - \frac{Q_1(y,t)}{q_2}\right] = O_p\left(\max_{1\le i\le n} m_i\inv n^{1/2}\right).\]
Under the assumption $\lim_{n\ra \infty} \max_{1\le i\le n} m_i\inv n^{1/2} = 0$ and by Slutsky's theorem, it suffices to show the asymptotic normality of $Q_1(y,t)/q_2$.
By the CLT, as $n\ra \infty$,
\[\sqrt n [Q_1(y,t) - q_1] \cinD \cN\left\{0,\sigma_1^2\theta'(t)^2\right\},\]
where $\sigma_1^2 = [F_{t,t}(y,y) - F_t(y)^2]$. By Slutsky's theorem, as $n\ra\infty$,
\[\sqrt n \left[\frac{Q_1(y,t)}{q_2} - \frac{q_1}{q_2}\right] \cinD \cN \left(0, \sigma_1^2\right].\]
Denote
$\beta_1 = \half\sigma^2(H')\pty f_t(y) + \half \sigma^2(K)\theta'(t)\inv \p2t [F_t(y)\theta'(t)]$ and $\beta_2 = \half\sigma^2(K)\theta^{(3)}(t)$. 
By Lemma \ref{lem:BiasVar},
\[
\lim_{n\ra \infty} \sqrt n \left[\frac{q_1}{q_2} - F_t(y)\right] = \beta_1 - \beta_2 F_t(y)/\theta'(t).
\]
By Slutsky's theorem, as $n\ra\infty$,
\[\sqrt n \left[\frac{Q_1(y,t)}{q_2}- F_t(y)\right] \cinD \cN \left\{ \beta_1 - \beta_2 F_t(y)/\theta'(t), \sigma_1^2\right\},\]
and hence as $n,m_i\ra\infty$,
\[\sqrt n \left\{\wt F_t(y)- F_t(y) \right\} \cinD \cN \left\{ \beta_1 - \beta_2 F_t(y)/\theta'(t), \sigma_1^2\right\}.\]
\epf
\fi
\bpf[Proof of Theorem \ref{thm:ANofF}]
With the optimal bandwidths of the order
$h_Y\sim n^{-1/4}$ and $h_T \sim n^{-1/4}$ and
by  Lemma \ref{lem:riem}, as $n,m_i\ra \infty$,
\begin{align*}
\sqrt n\left[\bpmt\wt D_1(y,t)\\ \wt D_2(y,t)\epmt - \bpmt Q_3(y,t)/q_2 - Q_1(y,t)q_4/q_2^2\\ Q_5(y,t)/q_2\epmt\right]
&= \bpmt O_p\left(\max_{1\le i\le n} m_i\inv n^{3/4}\right) \\ O_p\left(\max_{1\le i\le n} m_i\inv n^{3/4}\right) \epmt.
\end{align*}
Under the assumption $\lim_{n\ra \infty} \max_{1\le i\le n} m_i\inv n^{3/4} = 0$,  it suffices to show the asymptotic normality of $\left(Q_3(y,t)/q_2 - Q_1(y,t)q_4/q_2^2, Q_5(y,t)/q_2\right)^\tp$.
By the CLT, as $n\ra \infty$,
\[\sqrt n \left[\bpmt Q_1(y,t)\\ Q_3(y,t)\\ Q_5(y,t)\epmt - \bpmt q_1\\ q_3\\ q_5\epmt \right] \cinD \cN\left(\bm 0, \bm\Sigma\right), \text{ where }
\bm\Sigma = \lim_{n\ra\infty} \bpmt v_1 & v_{13} & v_{15}\\ v_{13} & v_3 & v_{35}\\ v_{15} & v_{35} & v_5\epmt.\]
Note that $\lim_{n\ra\infty} q_4/q_2^2 = \theta''(t)\theta'(t)^{-2}$ and $\lim_{n\ra\infty} 1/q_2 = \theta'(t)\inv$. 
Denote
\begin{gather*}
\bm\beta = \lim_{n\ra \infty} \sqrt n \left[ \bpmt q_1q_4/q_2^2 \\ q_3/q_2 \\ q_5/q_2 \epmt  - \bpmt F_t(y)\theta''(t)/\theta'(t) \\ \ptt[F_t(y)\theta'(t)]/\theta'(t) \\ f_t(y) \epmt \right].
\end{gather*}
By Slutsky's theorem, as $n\ra\infty$,
\[\sqrt n \left[\bpmt Q_1(y,t)q_4/q_2^2\\ Q_3(y,t)/q_2\\ Q_5(y,t)/q_2 \epmt - \bpmt F_t(y)\theta''(t)/\theta'(t) \\ \ptt[F_t(y)\theta'(t)]/\theta'(t) \\ f_t(y) \epmt\right] \cinD \cN\left(\bm \beta, \mbf A_1 \bm\Sigma \mbf A_1\right),\]
where $\mbf A_1 = \diag\{\theta''(t)\theta'(t)^{-2}, \theta'(t)\inv, \theta'(t)\inv\}$. 
By the continuous mapping theorem, as $n\ra\infty$,
\[
\sqrt n \left[\bpmt Q_3(y,t)/q_2 - Q_1(y,t)q_4/q_2^2 \\ Q_5(y,t)/q_2 \epmt - \bpmt \ptt F_t(y) \\ f_t(y) \epmt\right] \cinD \cN\left(\mbf A_2\bm \beta, \mbf A_2\mbf A_1 \bm\Sigma \mbf A_1\mbf A_2^\tp \right),
\]
where $\mbf A_2 = \bpmt -1 & 1 & 0\\ 0 & 0 & 1\epmt$.
Thus as $n,m_i\ra\infty$,
\[\sqrt n \left[\bpmt \wt D_1(y,t)\\ \wt D_2(y,t)\epmt - \bpmt D_1(y,t)\\ D_2(y,t)\epmt \right] \cinD \cN\left(\mbf A_2\bm \beta, \mbf A_2\mbf A_1 \bm\Sigma \mbf A_1\mbf A_2^\tp \right),\]
where by Lemma \ref{lem:BiasVar},
\begin{gather*}
\mbf A_2\bm\beta = \bpmt \half\sigma^2(H') \frac{\pt^2}{\pt t\pt y}f_t(y) + \half\sigma^2(K) \left[\frac{\pt^3}{\pt t^3}F_t(y)  + 2\ptt\left(\frac{\theta''(t) \ptt F_t(y)}{\theta'(t)}\right)\right]\\
\half\sigma^2(K) \left[\frac{\pt^2}{\pt y^2}f_t(y) + \p2t f_t(y) + 2\frac{\theta''(t) \ptt f_t(y)}{\theta'(t)}\right]\epmt,\\
\mbf A_2\mbf A_2\bm\Sigma \mbf A_1\mbf A_2^\tp = \bpmt \frac{\pt^2}{\pt s\pt s'}F_{t,t}(y,y) - \left[\ptt F_t(y)\right]^2 & \frac{\pt^2}{\pt s\pt z'}F_{t,t}(y,y) - f_t(y)\ptt F_t(y) \\
\frac{\pt^2}{\pt s\pt z'}F_{t,t}(y,y) - f_t(y)\ptt F_t(y)  & f_{t,t}(y,y) - f_t(y)^2\epmt.
\end{gather*}
For a given curve $y(t)$, the two components of the rank derivative and the corresponding estimates are $C_1(t) = D_1(y(t),t)$, $C_2(t) = D_2(y(t),t)y'(t)$, and $\wt C_1(t) = \wt D_1(y(t),t)$, $\wt C_2(t) = \wt D_2(y(t),t)y'(t)$, respectively.  Again by Slutsky's theorem, the proof is complete.
\epf

\references
\end{document}